\documentclass[final,1p,times]{elsarticle}

\usepackage{amssymb}
\usepackage{amsmath,amssymb,amsfonts}
\usepackage{multirow}
\usepackage{graphicx}
\usepackage{textcomp}
\usepackage{amsthm}
\usepackage{setspace}

\usepackage[noend]{algpseudocode}
\usepackage{algorithmicx,algorithm}
\usepackage{subfigure}
\usepackage{multicol}
\usepackage{color}
\usepackage{xspace}
\usepackage{pdfwidgets}

\begin{document}
\newtheorem{theorem}{Theorem}
\newtheorem{lemma}{Lemma}
\newtheorem{corollary}{Corollary}
\newtheorem{definition}{Definition}
\newtheorem{proposition}{Proposition}
\newtheorem{remark}{Remark}
\newtheorem{claim}{Claim}
\renewcommand{\algorithmicrequire}{\textbf{Input:}}
\renewcommand{\algorithmicensure}{\textbf{Output:}}
\bibliographystyle{plain}


\begin{frontmatter}



\title{A Random Algorithm for Profit Maximization with Multiple Adoptions in Online Social Networks}


\author[ouc]{Tiantian Chen}
\author[ouc]{Bin Liu\corref{cor1}}
\ead{binliu@ouc.edu.cn}
\author[ouc]{Wenjing Liu}
\author[ouc]{Qizhi Fang}
\author[td]{Jing Yuan}
\author[td]{Weili Wu}


\cortext[cor1]{Corresponding author}
\address[ouc]{School of Mathematical Sciences, Ocean University of China, Qingdao 266100, China}
\address[td]{Department of Computer Science, The University of Texas at Dallas, Richardson, TX 75080, USA}


\begin{abstract}
\label{abstract}
Online social networks have been one of the most effective platforms for marketing and advertising. Through ``word of mouth'' effects, information or product adoption could spread from some influential individuals to millions of users in social networks. Given a social network $G$ and a constant $k$, the influence maximization problem seeks for $k$ nodes in $G$ that can influence the largest number of nodes. This problem has found important applications, and a large amount of works have been devoted to identifying the few most influential users. But most of existing works only focus on the diffusion of a single idea or product in social networks. However, in reality, one company may produce multiple kinds of products and one user may also have multiple adoptions. 

Given multiple kinds of different products with different activation costs and profits, it is crucial for the company to distribute the limited budget among multiple products in order to achieve profit maximization. \emph{Profit Maximization with Multiple Adoptions} (PM$^{2}$A) problem aims to seek for a seed set within the budget to maximize the overall profit. In this paper, a \emph{Randomized Modified Greedy} (\emph{RMG})  algorithm based on the Reverse Inf\mbox{}luence Sampling (RIS) technique is presented for the PM$^{2}$A problem, which could achieve a $(1-1/e-\varepsilon)$-approximate solution with high probability. Compared with the algorithm proposed in \cite{Zhang16}  that achieves a $\frac{1}{2}(1-1/e^{2})$-approximate solution, our algorithm provides a better performance ratio which is also the best performance ratio of the PM$^{2}$A problem. Comprehensive experiments on three  real-world social networks are conducted, and the results demonstrate that our \emph{RMG} algorithm outperforms the algorithm proposed in \cite{Zhang16} and other heuristics in terms of profit maximization, and could better allocate the budget.
\end{abstract}


\end{frontmatter}

\section{Introduction}
\label{intro}
With the increasing popularity of online social and information networks such as Facebook, Twitter, LinkedIn, etc., many researchers have studied the diffusion phenomenon in social networks, including the diffusion of news, ideas, innovations, the adoption of new products, etc. Such diffusion is driven by the influence propagation throughout the social networks. One topic that has been extensively studied is the \emph{Influence Maximization} (IM) problem \cite{Borodin, Chen10, Chen09, Kempe03}. The goal of the IM problem is to find a small subset of influential nodes such that they can attract the largest number of members in a social network, according to an influence propagation model. The IM problem has been mainly studied under two classical influence propagation models: \emph{Independent Cascade} (IC) model \cite{Goldenberg1, Goldenberg2} and \emph{Linear Threshold} (LT) model \cite{Granovetter, Schelling}. Both IC model and LT model are probabilistic models that characterize how the influence is propagated in the social network starting from an initial set of seed nodes.

The objective functions of the IM related problems are usually complicated to compute due to the randomness of the probabilistic diffusion models. In fact, computing the expected influence for a given seed set is $\#$P-hard \cite{Chen10}. To address this issue, Kempe et al.\ \cite{Kempe03} use Monte-Carlo simulations to estimate the expected influence in each iteration of a natural greedy algorithm, which incurs significant computational overheads. Such inefficiency has motivated a large amount of research on the IM problem in the past decade \cite{Chen09,Chen102, Goyal, Jung, Kempe03, Kempe05, Leskovec}. However, most of these methods either trade performance guarantees for practical efficiency, or vice versa. There are exceptions like \emph{TIM}/\emph{TIM$^{+}$} \cite{Tang14} and \emph{IMM} \cite{Tang15}, which are scalable methods with performance guarantee for the IM problem. They utilized a novel Reverse Influence Sampling (RIS) technique introduced by Borgs et al.\ \cite{Borgs14} and obtained $(1-1/e-\varepsilon)$-approximate solutions with high probability .

Most existing works focus on the IM problem with a single diffusion, i.e., only one product is considered. In reality, however, one company may produce multiple products and people may purchase multiple kinds of products at one time. For example, Apple sells iPhone, iPad, Macbook, etc. Many people have iPhone and iPad at the same time, and plenty of people own both laptop and desktop. Therefore, given a limited budget and some kinds of different items with different activation costs and profits, a crucial question is: \emph{how to allocate the budget to maximize the overall profit?} Recently, Zhang et al.\ \cite{Zhang16} formulated a \emph{Profit Maximization with Multiple Adoption} (PM$^{2}$A) problem, which seeks for a seed set within the limited budget to massively influence customers and achieves the goal of profit maximization. And a $\frac{1}{2}(1-1/e^2)$-approximation algorithm was presented by Zhang et al. Moreover, they proposed another algorithm, called \emph{PMIS}, and stated that \emph{PMIS} could produce a solution within a factor of $\alpha \cdot (1-1/e)$, where $\alpha$ may be made arbitrarily close to $1$. However, $\alpha$ is obtained by using CPLEX to solve the \emph{Multiple-Chioce Knapsack} problem.

In this paper, we present an efficient algorithm called \emph{Randomized Modified Greedy} (\emph{RMG}) algorithm for the PM$^{2}$A problem. The \emph{RMG} algorithm based on the RIS technique returns a $(1-1/e-\varepsilon)$-approximate solution with at least $1-(nq)^{-l}-3qn^{-l'}$ probability, where $q$ is the number of products, $n$ is the number of nodes in the social network, and $\varepsilon,l,l'>0$ are constants. The \emph{RMG} algorithm can be implemented with tunable parameters and it is flexible for balancing the running time and the accuracy. Experimental results show that the \emph{RMG} algorithm not only produces high quality seed sets but also takes much less time than the greedy algorithm with Monte-Carlo simulations.

\medskip

The contributions of this paper are summarized as follows:
\vspace{-0.1in}
\begin{itemize}
\setlength{\itemsep}{0pt}
\setlength{\parsep}{0pt}
\setlength{\parskip}{0pt}
       \item  We present a \emph{Randomized Modified Greedy} (\emph{RMG}) algorithm  for the PM$^{2}$A problem that achieves a $(1-1/e-\varepsilon)$-approximation ratio with high probability, which significantly improves upon prior works in terms of performance guarantee and is also the best performance ratio of the PM$^{2}$A problem even for one product.
       \item We extend the RIS technique to accommodate the profit estimation over multiple products, and an \textit{OPT} estimation scheme is designed to speed up the sampling process, where \textit{OPT} is the optimum of the PM$^{2}$A problem.
       \item We conduct comprehensive experiments on real-world social networks that verify the superiority of \emph{RMG} in providing an effective budget distribution over multiple products.
\end{itemize}
\vspace{-0.05in}

\noindent
\textbf{Related works.}\ The Influence Maximization (IM) problem has been studied intensively in the past decade. Kempe et al. \cite{Kempe03} first formulated IM as a combinatorial optimization problem and presented a general greedy algorithm that yielded a $(1-1/e-\varepsilon)$-approximation for all diffusion models considered. Later, a large amount of works aimed to improve the efficiency and scalability of the seed selection algorithm. However, they were either heuristics without performance guarantees \cite{Chen10, Chen09} or prohibitively slow on billion-scale networks \cite{Goyal, Leskovec}. Borgs et al.\ \cite{Borgs14} made a theoretical breakthrough with the RIS technique that guaranteed a $(1-1/e-\varepsilon)$-approximation and significantly reduced the expected running time. Subsequently, \cite{Tang14, Tang15, Nguyen16} proposed algorithms that were very efficient even on large networks with millions of nodes and billions of edges.

Recently, research on influence or profit maximization with a limited budget has also emerged. Nguyen et al.\ \cite{Nguyen13} consider the budgeted influence maximization problem in which each node can have an arbitrary cost, but they mainly focus on the diffusion of a particular kind of product, which is different from our work. Multiple products are considered in \cite{Du14}, where seperate budget constraints are made on different products and their diffusions are separated as well. However, all of the products share an overall activation budget in the PM$^{2}$A problem. The PM$^{2}$A problem is proposed by Zhang et al.\cite{Zhang16} and a $\frac{1}{2}(1-1/e^{2})$-approximation algorithm is discussed under the  \emph{Multiple Thresholds} (MT) model which is an extension of LT model. They proposed another algorithm, called \emph{PMIS}, and stated that \emph{PMIS} could produce a solution within a factor of $\alpha \cdot (1-1/e)$, where $\alpha$ may be made arbitrarily close to $1$. However, $\alpha$ is obtained by using CPLEX to solve the \emph{Multiple-Chioce Knapsack} problem.\cite{Hung17} investigates the cost-aware targeted viral marketing problem in which only one product spreads in the network and each node has its own selecting cost and benefit, and aims to find a seed set with total cost no more than a budget to maximize the expected total profit. The PM$^{2}$A problem is actually a special case of the problem considered in \cite{Hung17}, but they only propose a $(1-1/\sqrt{e}-\varepsilon)$-approximation algorithm. In this paper, we present a $(1-1/e-\varepsilon)$-approximation algorithm based on the RIS technique under the IC model, which significantly improves the solution of the PM$^{2}$A problem.

\noindent
\textbf{Organization.} \ The rest of the paper is organized as follows.  In Section \ref{Preliminary}, we introduce the diffusion model and the definition of the PM$^{2}$A problem. Key ideas of solving the PM$^{2}$A problem and the framework of the algorithm are presented in Section \ref{Key ideas}. Section \ref{Alg} is dedicated to the \emph{RMG} algorithm along with the analysis of its performance ratio and time complexity. Section \ref{Experiments} shows our experimental results and Section \ref{Conclusion} concludes the paper.

\section{Problem Formulation}
\label{Preliminary}

This work aims to design a marketing strategy for allocating the budget among multiple products in a social network. In this section, we present the diffusion model and give the definition of the \emph{Profit Maximization with Multiple Adoptions} (PM$^{2}$A) problem.

A \emph{social network} is usually represented as a digraph $G=(V,E)$ with nodes in $V$ representing users and edges in $E$ representing relationships between users, where $|V|=n,|E|=m$. Assume that each directed edge $e$ in $G$ is associated with a propagation probability $p(e)\in [0,1]$.

\subsection{Diffusion Model}
There are many diffusion models studied in the literature. The diffusion model considered in this paper is the \emph{Independent Cascade} (IC) model, investigated in the context of marketing by Goldenberg et al.\ \cite{Goldenberg1, Goldenberg2}. Given a social network $G$, the IC model considers a timestamped influence propagation process as follows:

\textrm{1.} At timestamp $1$, we activate a selected node set $S\subseteq V$, and set all of the remaining nodes inactive.

\textrm{2.} If a node $v$ is first activated at timestamp $t$, then for each directed edge $e$ pointing from $v$ to an inactive node $u$, $v$ has a probability $p(e)$ to activate $u$ at timestamp $t+1$. After timestamp $t+1$, $v$ has no chance to activate any node.

\textrm{3.} Once a node becomes active, it remains active in the following timestamps.

Let $I(S)$ be the number of nodes that are activated when the above diffusion process terminates. Refer to $S$ as the \emph{seed set}, and $I(S)$ as the \emph{spread} of $S$. Let $\sigma(S)$ be the \emph{expected spread} of $S$, that is $\sigma(S)=\mathbb{E}[I(S)]$.

\subsection{Problem Definition}
In reality, one company may produce multiple different products and one user may purchase multiple kinds of products at one time. Therefore, given a limited budget and some kinds of different  items with different activation costs and profits, it is crucial for the company to wisely allocate the budget to maximize the overall profit. This problem is called the Profit Maximization with Multiple Adoptions (PM$^{2}$A) problem, which was introduced by Zhang et al. \cite{Zhang16}.

\begin{definition}[Profit Maximization with Multiple Adoptions (PM$^{2}$A)]
 Given a social network $G=(V,E)$ with propagation probability $p:E\rightarrow [0,1]$, suppose there are $q$ different kinds of products spreading independently in $G$ and each node can adopt multiple products. For $i=1,2,\ldots,q$, let $c_{i}$ be the cost of initially activating a node to adopt product $i$, and $p_{i}$ be the profit obtained when a node is activated to adopt product $i$, where $c_{i}, p_{i}>0$. The $PM^{2}A$ problem asks to identify a seed set for each product respectively with overall activation cost at most $B$ such that the expected total profit is maximized.
 \end{definition}

Obviously, the Influence Maximization (IM) problem is a special case of the PM$^{2}$A problem when $q=1$. Since the IM problem is NP-hard for the IC model \cite{Kempe03} and cannot be approximated better than $1-1/e$ unless P=NP \cite{Kempe15}, we have the following result.
\begin{claim}
The PM$^{2}$A problem is NP-hard for the IC model, and for any $\varepsilon>0$, it cannot be approximated in polynomial time within a ratio of $(1-1/e+\varepsilon)$ unless P = NP.
\end{claim}

Since each node may adopt multiple products, the PM$^{2}$A problem can be characterized as $q$ independent diffusion processes under a common budget constraint in $G$. The most crucial point here is how to allocate the budget among the multiple products. To address this issue, we shall give another version of the PM$^{2}$A problem in the next section. 
\section{Key Ideas for Solving PM$^{2}$A Problem}
\label{Key ideas}
In this section, we transfer the PM$^{2}$A problem into an equivalent problem, called PM-$\widetilde{G}$, and present the framework for solving the PM-$\widetilde{G}$ problem.

\subsection{Reformulation of the Problem}
\begin{definition}[ $q$-component copy graph $\widetilde{G}$ ]
Given a social network  $G=(V,E)$ with propagation probability $p:E\rightarrow [0,1]$, let $\widetilde{G}=(\widetilde{V},\widetilde{E})=G^{(1)}\cup G^{(2)}\cup \cdots \cup G^{(q)}$ be a graph composed of $q$ components, each of which (denoted by $G^{(i)}=(V^{(i)},E^{(i)})$) is a copy of $G$. For any node $u\in V$, let $u^{(i)}$ be the copy node of $u$ in $G^{(i)}$. For different copy nodes $u^{(i)}$ and  $u^{(j)}$ of $u$ in different components $G^{(i)}$ and $G^{(j)}$, $i \neq j$, they are regarded as different nodes in the $\widetilde{G}$.
Then for any node set $S\subseteq \widetilde{V}$, $S$ can be described as $S=S^{(1)}\cup S^{(2)}\cup \ldots\cup S^{(q)}$, where $S^{(i)}\subseteq V^{(i)}$, $i=1,2,\ldots,q$. 
\end{definition}

\begin{figure}[htbp]
\vspace{-0.5cm}  
\setlength{\abovecaptionskip}{-0.2cm}  
\setlength{\belowcaptionskip}{-0.4cm}   
\centerline{\includegraphics[width=1.0\textwidth]{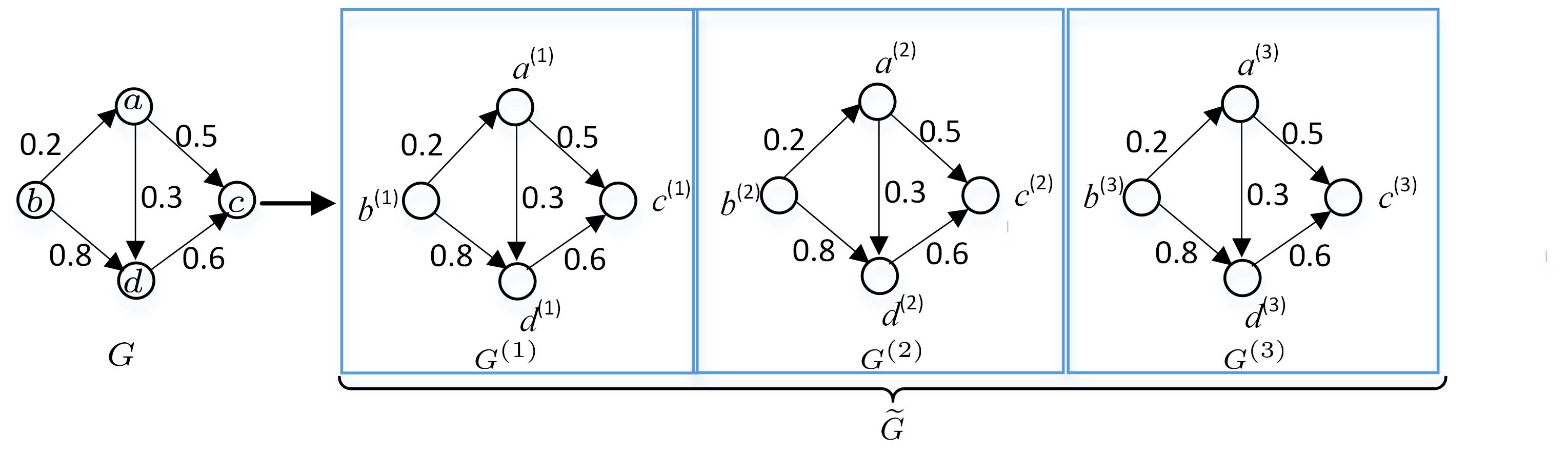}}
  \caption{An example of $q$-component copy graph $\widetilde{G}$, $q=3$.}
\label{fig2}
\end{figure}

Thus,  $q$ different products spreading in $G$ are transformed to a single product spreading in $\widetilde{G}$, subject to the condition that both the activation cost and the profit are different in different components:
\vspace{-0.1in}
\begin{itemize}
\setlength{\itemsep}{0pt}
\setlength{\parsep}{0pt}
\setlength{\parskip}{0pt}
\item{Cost in $G^{(i)}$:} the cost of initially activating a node in component $G^{(i)}$ to adopt the product is $c_{i}$.
\item{Profit in $G^{(i)}$:} the profit when a node in component $G^{(i)}$ is activated to adopt the product is $p_{i}$.
\end{itemize}
\vspace{-0.1in}

Since different components do not connect each other, for a seed set $S=S^{(1)}\cup S^{(2)}\cup \ldots\cup S^{(q)}$, $S^{(i)}$ can only influence the nodes in $G^{(i)}$. For $i=1,2,\ldots,q$, denote by $\sigma(S^{(i)})$  the expected spread of $S^{(i)}$, and $\rho(S^{(i)})$ the excepted profit gained by initially activating the nodes in $S^{(i)}$, i.e., $\rho(S^{(i)})=p_{i}\cdot \sigma(S^{(i)})$. Let $\rho(S)$ be the expected total profit gained by initially activating all the nodes in $S$, then $\rho(S)=\sum_{i=1}^{q}{p_{i}\cdot \sigma(S^{(i)})}.$
Let $c(S)$ be the activation cost of $S$, that is, $c(S)=\sum_{i=1}^{q}{c_{i}|S^{(i)}|}.$

\begin{definition}[Profit Maximization on $\widetilde{G}$ (PM-$\widetilde{G}$)]
 The PM-$\widetilde{G}$ problem asks for a seed set $S\subseteq \widetilde{V}$  with the activation cost at most $B$ such that the expected total profit is maximized:
\begin{equation}\label{eqn2}
\setlength{\abovedisplayskip}{1pt}
\setlength{\belowdisplayskip}{0pt}
 \begin{aligned}
  &\max\ \,\, \rho(S)=\sum_{i=1}^{q}{p_{i}\cdot \sigma(S^{(i)})}\\
  &s.t. \quad \sum_{i=1}^{q}{c_{i}|S^{(i)}|}\le B
 \end{aligned}
\end{equation}
\end{definition}

It is easy to see that the PM$^{2}$A problem is equivalent to the PM-$\widetilde{G}$ problem.

Let $\Omega=\{S\subseteq \widetilde{V}\mid c(S)\le B\}$, that is, $\Omega$ is the feasible set of the PM-$\widetilde{G}$ problem. Let $S^{*}$ be the optimal solution of the PM-$\widetilde{G}$ problem, and $\textit{OPT}=\rho(S^{*})$ be the optimum.

Since $\rho(S)=\sum_{i=1}^{q}{p_{i}\cdot \sigma(S^{(i)})}$ and $\sigma(S^{(i)})$ has been proved to be nondecreasing as well as submodular \cite{Kempe03}, the following result holds.
\begin{proposition}
The profit function $\rho(\cdot)$ in the PM-$\widetilde{G}$ problem is nonnegative, nondecreasing and submodular.
\end{proposition}

Given multiple items with different costs and profits, and a limited budget, one may resort to algorithms of the classical knapsack problem, such as greedy algorithm and dynamic programming.  However, those methods cannot perform well here, since the profit function $\rho(\cdot)$ in the PM-$\widetilde{G}$ problem is submodular rather than linear. And selecting any node from the candidate node set may influence the marginal gain of choosing the next seed, not like the static weight associated with each item in the knapsack problem. All  the facts make the PM-$\widetilde{G}$ problem difficult.

For the problem of maximizing a nonnegative, nondecreasing submodular set function $f(\cdot)$ subject to a knapsack constraint, Sviridenko\cite{Sviridenko04} proposed a modified greedy algorithm which guarantees a $(1-1/e)$-approximation ratio and is based on a value oracle model for $f(\cdot)$. That is, for a given set $S$, the algorithm can query an oracle to find its value $f(S)$. But our task in this work is to solve the PM-$\widetilde{G}$ problem without using the value oracle model, and it is accompanied by the difficulty of computing $\rho(\cdot)$, because the computation of $\sigma(\cdot)$ has been shown to be $\#$P-hard\cite{Chen10}.
\vspace{-0.05in}
\subsection{Framework for Solving PM-$\widetilde{G}$ Problem}
Before we give the algorithm for solving the PM-$\widetilde{G}$ problem in detail, we first describe the main idea and framework of the algorithm. 

To tackle intractability of the computation of $\rho(\cdot)$, we try to obtain an estimate $\hat{\rho}(\cdot)$ of $\rho(\cdot)$ with a small error with high probability, where $\hat{\rho}(\cdot)$ can be computed in polynomial time. Then we substitute $\hat{\rho}(S)$ for $\rho(S)$ to translate the original PM-$\widetilde{G}$ problem into maximizing $\hat{\rho}(S)$ with the budget constraint. Using the modified greedy algorithm \cite{Sviridenko04}, a $(1-1/e)$-approximate solution $S_{\mathcal{A}}$ for the problem of $\max_{S\in \Omega}\hat{\rho}(S)$ is obtained,  which can be proved to be a $(1-1/e-\varepsilon)$-approximate solution for the PM-$\widetilde{G}$ problem with high probability.

In the estimation of $\rho(S)$, we utilize the Reverse Influence Sampling (RIS) technique introduced by Borgs et al. \cite{Borgs14}, which significantly improves the time complexity of the algorithm for the IM  problem.

\begin{figure}[htbp]
\vspace{-0.2cm}  
\setlength{\abovecaptionskip}{0.1cm}   
\setlength{\belowcaptionskip}{-0.3cm}
\centerline{\includegraphics[width=0.8\textwidth]{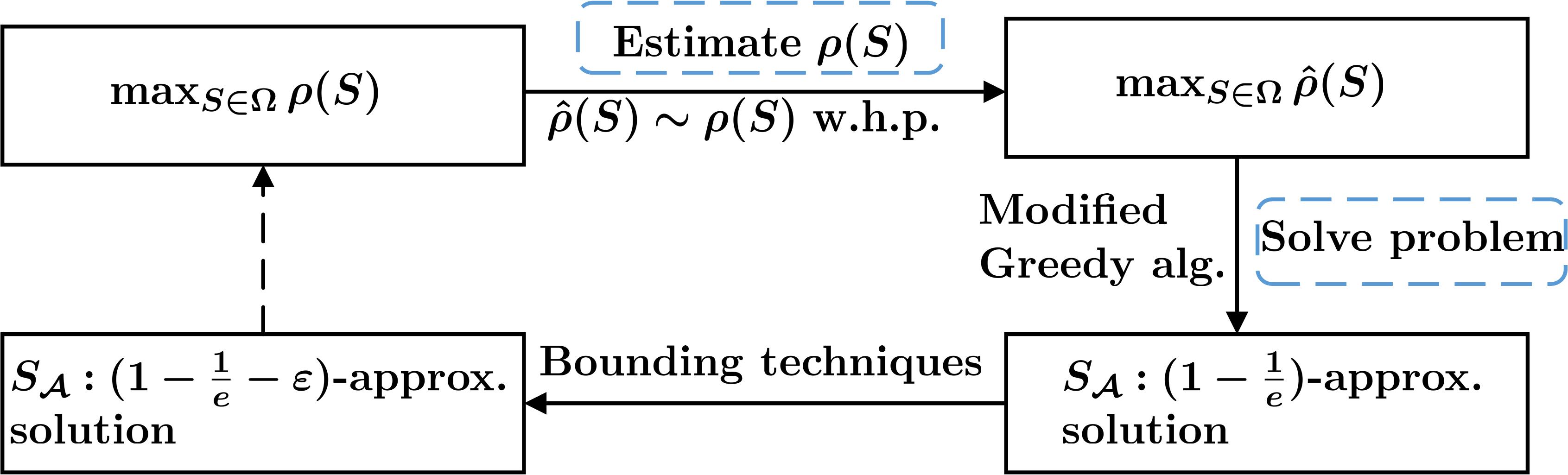}}
  \caption{Overview of algorithms.}
\label{fig1}
\end{figure}

In summary, the PM-$\widetilde{G}$ problem can be solved by the following steps.
\vspace{-0.1in}
\begin{itemize}
\setlength{\itemsep}{0pt}
\setlength{\parsep}{0pt}
\setlength{\parskip}{0pt}
\item \textbf{Estimate} $\boldsymbol{\rho(S):}$ Use the RIS technique to gain an estimation $\hat{\rho}(\cdot)$ of $\rho(\cdot)$ such that for any $S\in \Omega$, $|\hat{\rho}(S)-\rho(S)| < \frac{\varepsilon}{2}\cdot OPT$ holds with high probability, where $0<\varepsilon<1$.
\item \textbf{Solve problem} $\boldsymbol{ \max_{S\in \Omega}\hat{\rho}(S):}$ Prove $\hat{\rho}(S)$ is nonnegative, nondecreasing and submodular, then use the Modified Greedy algorithm to solve the problem $\max_{S\in \Omega}\hat{\rho}(S)$.
    Let $S_{\mathcal{A}}$ be the solution returned by the algorithm, then we can show that $S_{\mathcal{A}}$ is a $(1-1/e-\varepsilon)$-approximate solution for the PM-$\widetilde{G}$ problem with high probability (w.h.p.).
\end{itemize}
\vspace{-0.1in}
The framework for solving the PM-$\widetilde{G}$ problem is shown in Fig.\ref{fig1}.
\vspace{-0.15in}

\section{Algorithm and Its Analysis}
\label{Alg}
In this section, the \emph{Randomized Modified Greedy} (\emph{RMG}) algorithm for the PM-$\widetilde{G}$ problem is presented, which can achieve a $(1-1/e-\varepsilon)$-approximation ratio with high probability. Before introducing the algorithm, we list some notations for convenience. Let $p_{min}=\min_{1\le i \le q}\{p_{i}\}$, $p_{max}=\max_{1\le i \le q}\{p_{i}\}$, $c_{min}=\min_{1\le i \le q}\{c_{i}\}$ and $k_{i}=\lfloor B/c_{i}\rfloor$. Let $k^{*}=\lfloor B/c_{min}\rfloor$, i.e. the maximum number of seed nodes that can be chosen. As the budget $B$ is often limited, the size of the seed set can not be too large. Thus in the following, we assume that $k^{*}\le \lfloor nq/2\rfloor$.

\subsection{Estimation of $\rho(S)$}
Now we are in the position to give the estimation of $\rho(S)$. In this work, Reverse Influence Sampling (RIS) technique is used, which captures the influence landscape of the social network through generating a set of \emph{Random Reverse Reachable (RR) sets}\cite{Tang14}.

\noindent
\textbf{Random Reverse Reachable (RR) set.}\ Given a social network $\widetilde{G}=(\widetilde{V},\widetilde{E})$ with propagation probability $p:E\rightarrow [0,1]$, a random reverse reachable (RR) set is generated by 1) generating a sample graph  $g$ from $\widetilde{G}$ by removing each edge $e$ in $\widetilde{G}$ with $1-p(e)$ probability 2) selecting a node $v$ from $g$ uniformly at random 3) returning $R$ as the set of nodes that can reach $v$ in $g$.

\begin{figure}[htbp]
\vspace{-0.4cm}  
\setlength{\abovecaptionskip}{-0.1cm}   
\setlength{\belowcaptionskip}{-0.3cm} 
\centerline{\includegraphics[width=1.0\textwidth]{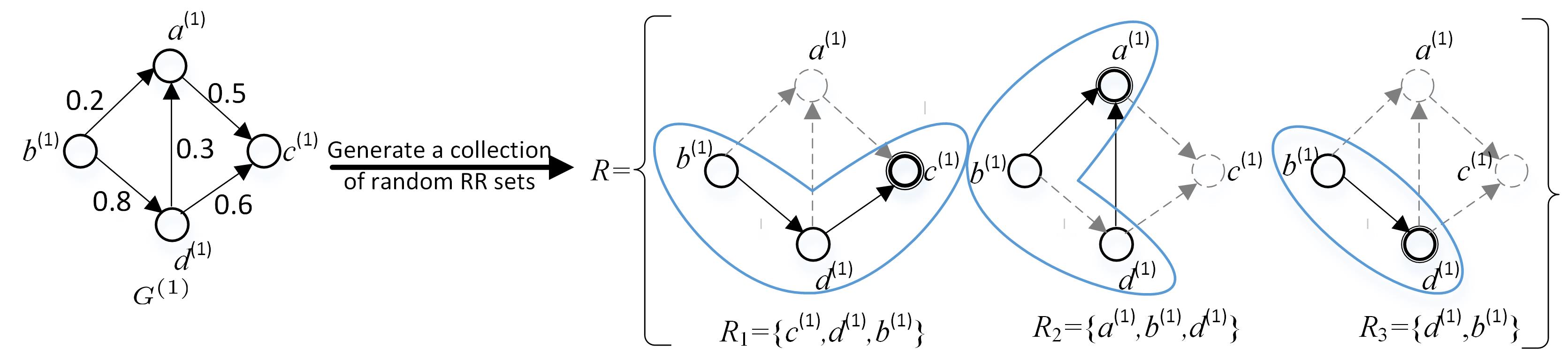}}
  \caption{An example of generating random RR sets under the IC model. Three random RR sets $R_{1}, R_{2}$ and $R_{3}$ are generated for three nodes $c^{(1)}, a^{(1)}$ and $d^{(1)}$, respectively.}
\label{fig2}
\end{figure}

Intuitively, if a node $u$ appears in an RR set generated for another node $v$, then $u$ can reach $v$ through a certain path in $\widetilde{G}$. Thus, a propagation process from a seed set containing $u$ should have a certain probability to activate $v$. The result of \cite{Borgs14} attests to this observation.

Using the reverse Breadth First Search (BFS) algorithm in \cite{Moore59}, we can
generate a set of random RR sets $\mathcal{R}=\{R_{1},R_{2},\ldots,R_{\theta}\}$. Given a node set $S\subseteq \widetilde{V}$,  we say that $S$ covers an RR set $R_{j}$ if and only if $S\cap R_{j}\neq \emptyset$. Define $F_{\mathcal{R}}(S)$ as the fraction of RR sets in $\mathcal{R}$ covered by $S$, that is
\[
\setlength{\abovedisplayskip}{1pt}
\setlength{\belowdisplayskip}{1pt}
F_{\mathcal{R}}(S)=\frac{|\{R_{j}\in \mathcal{R}, 1\leq j \leq \theta \mid S\cap R_{j}\neq \emptyset\}|}{\theta}.
\]

Recall that $S=S^{(1)}\cup S^{(2)}\cup \cdots \cup S^{(q)}\subseteq \widetilde{V}$, it is clear that $F_{\mathcal{R}}(S)=\sum_{i=1}^{q}F_{\mathcal{R}}(S^{(i)})$.
Based on the results in Tang et al.\ \cite{Tang14}, we can obtain that for any $S^{(i)}\subseteq V^{(i)}$, $i=1,2,\ldots,q$, the expected value of $nq\cdot F_{\mathcal{R}}(S^{(i)})$ equals the expected spread of $S^{(i)}$ in $\widetilde{G}$. This implies to the following lemma.
\begin{lemma}\label{lem1}
For any node set $S^{(i)}\subseteq V^{(i)}$, $\mathbb{E}[nq\cdot F_{\mathcal{R}}(S^{(i)})]=\sigma(S^{(i)})$, $i=1,2,\ldots,q$.
\end{lemma}

Denote $\hat{\sigma}(S^{(i)})=nq\cdot F_{\mathcal{R}}(S^{(i)})$. Then according to Lemma \ref{lem1}, $\hat{\sigma}(S^{(i)})$ is an unbiased estimate of $\sigma(S^{(i)})$.
Define $\hat{\rho}(S^{(i)})=p_i\hat{\sigma}(S^{(i)})$ and for any $S\subseteq\widetilde{V}$, let $\hat{\rho}(S)=\sum_{i=1}^{q}\hat{\rho}(S^{(i)})=\sum_{i=1}^{q}{p_{i}\hat{\sigma}(S^{(i)})}$. Obviously, $\hat{\rho}(S^{(i)})$ and $\hat{\rho}(S)$ are unbiased estimates of $\rho(S^{(i)})$ and $\rho(S)$, respectively.

\begin{corollary}
For any $S\subseteq\widetilde{V}$, $\mathbb{E}[\hat{\rho}(S^{(i)})]=\rho(S^{(i)}), (i=1,2,\ldots,q)$, and $\mathbb{E}[\hat{\rho}(S)]=\rho(S)$.
\end{corollary}

For any $S\in \Omega$, we use the following Algorithms \ref{alg1} and \ref{alg2} to obtain the value of $\hat{\rho}(S)$. In Algorithm \ref{alg1}, we generate a set of $\theta$ random RR sets, denoted by $\mathcal{R}$. In Algorithm \ref{alg2}, we first identify the nodes in $S$ and partition them into $S^{(1)},S^{(2)},\ldots,S^{(q)}$,  where $S^{(i)}\subseteq V^{(i)}$ (Lines 2-6), then compute the fraction of RR sets in $\mathcal{R}$ covered by $S^{(i)}$, denoted by $F_{\mathcal{R}}(S^{(i)})$ (Lines 7-10). Summing up all the $q$ items, we obtain an estimate $\hat{\rho}(S)$.

\begin{algorithm}[t]
\caption{RR Sets Generation} 
\label{alg1}
\hspace*{0.02in} {\bf Input:} 
Graph $\widetilde{G}$ and a positive integer $\theta$.\\
\hspace*{0.02in} {\bf Output:} 
a set of $\theta$ random RR sets $\mathcal{R}$.
\begin{algorithmic}[1]
\State Initialize: $\mathcal{R}=\emptyset$;  
\State Use the reverse Breadth First Search algorithm to generate $\theta$ random RR sets and insert them into $\mathcal{R}$;  
\State Initialize: $B=\emptyset$, $\sigma(\emptyset)=0$;  
\State \Return $\mathcal{R}$.
\end{algorithmic}
\end{algorithm}

\begin{algorithm}[htbp]
\caption{Profit-Estimate} 
\label{alg2}
\hspace*{-0.02in} {\bf Input:} 
A seed set $S=\{\bar{v}_{1},\bar{v}_{2},\cdots,\bar{v}_{|S|}\} \in \Omega$, $\mathcal{R}=\{R_{1},R_{2},\cdots,R_{\theta}\}$, $0<\varepsilon <1$.\\
\hspace*{-0.02in} {\bf Output:} 
$\hat{\rho}(S)$ such that $|\hat{\rho}(S)-\rho(S)|<\frac{\varepsilon}{2}\cdot OPT$ with at least $1-(nq)^{-l}(k^{*})^{-1}/\binom{nq}{k^{*}}$ probability..
\begin{algorithmic}[1]
\State Initialize: $\hat{\rho}(S)=0$;  
\For{$i$ from $1$ to $q$}
¡¡¡¡\State $S^{(i)} \leftarrow \emptyset$;¡¡¡¡
     \For{$j$ from $1$ to $|S|$}
        \If{$\bar{v}_{j}\in V^{(i)}$}
        ¡¡¡¡\State $S^{(i)}=S^{(i)}\cup \{\bar{v}_{j}\}$;¡¡¡¡
        \EndIf
\EndFor
\EndFor
\For{$i$ from $1$ to $q$}
¡¡¡¡\State Initialize:$F_{\mathcal{R}}(S^{(i)})=0$;¡¡¡¡
     \For{$k$ from $1$ to $\theta$}
       ¡¡¡¡\State $F_{\mathcal{R}}(S^{(i)})=F_{\mathcal{R}}(S^{(i)})+\frac{\min \{|S^{(i)}\cap R_{k}|,1\}}{\theta}$;¡¡¡¡
\EndFor
\EndFor
\State \Return $\hat{\rho}(S)=\sum_{i=1}^{q}{nqp_{i}\cdot F_{\mathcal{R}}(S^{(i)})}$.
\end{algorithmic}
\end{algorithm}

By Chernoff bounds \cite{Chernoff}, we show that for any $S\in\Omega$, the result obtained by Algorithm \ref{alg2} is an accurate estimate of $\rho(S)$ with high probability, when $\theta$ is sufficiently large.

\begin{lemma}\label{lem2}
Suppose $\theta$ satisfies
\begin{equation}\label{eqn3}
\setlength{\abovedisplayskip}{1pt}
\setlength{\belowdisplayskip}{1pt}
\theta \ge (8q+2\varepsilon)nq^{2}p_{max}\cdot \frac{l\log (nq)+\log (2qk^{*})+\log \binom{nq}{k^{*}}}{\varepsilon^{2}\cdot \textit{OPT}}.
\end{equation}
Then for any set $S\in \Omega$, the following inequality holds with at least $1-(nq)^{-l}(k^{*})^{-1}/ \binom{nq}{k^{*}}$ probability:
\begin{equation}\label{eqn4}
\setlength{\abovedisplayskip}{1pt}
\setlength{\belowdisplayskip}{1pt}
|\hat{\rho}(S)-\rho(S)| < \frac{\varepsilon}{2}\cdot \textit{OPT},
\end{equation}
where $l>0$, $0<\varepsilon<1$ and $k^{*}=\lfloor B/c_{min}\rfloor$.
\end{lemma}

Since $\hat{\rho}(S)\sim \rho(S)$ with high probability and $\hat{\rho}(S)$ can be computed in polynomial time, we now turn to solve the following problem.
\vspace{-0.1in}
\begin{equation}\label{eqn7}
 \begin{aligned}
\setlength{\belowdisplayskip}{1pt}
  &\max\ \,\, \hat{\rho}(S)\\
  &s.t. \quad \sum_{i=1}^{q}{c_{i}|S^{(i)}|}\le B
 \end{aligned}
\end{equation}

\subsection{Modified Greedy Algorithm for the Problem $\max_{S\in \Omega}\hat{\rho}(S)$}

In this section, we provide a Modified Greedy algorithm for problem (\ref{eqn7}) which achieves a ($1-1/e$)-approximate solution $S_{\mathcal{A}}$. Then we show that $S_{\mathcal{A}}$ is a $(1-1/e-\varepsilon)$-approximate solution for the original PM-$\widetilde{G}$ problem with high probability. 

\begin{lemma}\label{lem4}
 $\hat{\rho}(\cdot)$ is nonnegative, nondecreasing and submodular.
\end{lemma}

Motivated by the design of the main algorithm in \cite{Sviridenko04}, we propose a Modified Greedy algorithm for problem (\ref{eqn7}). The sketch of the algorithm is as follows. We first enumerate all the feasible seed sets containing one or two nodes separately, to avoid the extreme situation that nodes with high profit and cost are not included in the solution (Lines 2-5). Then start with any feasible seed set consisting of three nodes, and greedily add node which does not destroy the feasibility of the set (Lines 6-18). Finally, output the maximum among the two cases (Line 19).

\begin{algorithm}[t]
\caption{Modified Greedy Algorithm} 
\label{alg3}
\hspace*{0.02in} {\bf Input:} 
Graph $\widetilde{G}$,  a budget $B$ and $0< \varepsilon < 1$.\\
\hspace*{0.02in} {\bf Output:} 
A $(1-1/e-\varepsilon)$-approximate solution for the PM-$\widetilde{G}$ problem, with at least $(1-(nq)^{-l})$-probability.
\begin{algorithmic}[1]
\State Initialize: $\overline{U}=\emptyset,\overline{S}=\emptyset,\widehat{V}=\widetilde{V}, \overline{V}=\emptyset$;  
\For{all $U \in \Omega,|U|=1$ or $2$}
¡¡¡¡\State $\hat{\rho}(U)\leftarrow$ \text{Profit-Estimate}$(\mathcal{R},U,\varepsilon)$;¡¡¡¡
¡¡¡¡\State Insert $U$ into $\overline{U}$;¡¡¡¡
\EndFor
\State $U^{*}=\arg \max_{U\in \overline{U}}\, \{\hat{\rho}(U)\}$;
\For{all $S_{0} \in \Omega,|S_{0}|=3$}
¡¡¡¡\State $S \leftarrow S_{0}$;¡¡¡¡
     \While{$c(S)\le B$}
     ¡¡¡¡\State $\widehat{V}=\widehat{V}\backslash S$;¡¡¡¡
     ¡¡¡¡\State $\hat{\rho}(S)\leftarrow$ \text{Profit-Estimate}$(\mathcal{R},S,\varepsilon)$;¡¡¡¡
          \For{all $v\in \widehat{V}$}
              \If{$c(S\cup \{v\})\le B$}
              ¡¡¡¡\State Insert $v$ into $\overline{V}$;¡¡¡¡
              ¡¡¡¡\State $\hat{\rho}(S \cup  \{v\})\leftarrow$ \text{Profit-Estimate}$(\mathcal{R},S\cup \{v\},\varepsilon)$;¡¡¡¡
              \EndIf
           \EndFor
      \State $v^{*}=\arg \max_{v\in \overline{V}}\, \{\frac{\hat{\rho}(S\cup \{v\})-\hat{\rho}(S)}{c(\{v\})}\}$;
      \State $S=S\cup\{v^{*}\}$;
      \EndWhile
  \State Insert $S$ into $\overline{S}$;   
\EndFor
\State $S_{0}^{*}=\arg \max_{S\in \overline{S}}\, \{\hat{\rho}(S)\}$;
\State \Return $S_{\mathcal{A}}=\text{arg\ max}\ \{\hat{\rho}(U^{*}),\hat{\rho}(S_{0}^{*})\}$.
\end{algorithmic}
\end{algorithm}

\begin{theorem}\label{thm1}
Given a graph $\widetilde{G}$, a positive number $B$, $0<\varepsilon<1$, $l>0$ and $\theta$ that satisfies inequality (\ref{eqn3}), Algorithm \ref{alg3} returns a $(1-1/e-\varepsilon)$-approximate solution for the PM-$\widetilde{G}$ problem with at least $1-(nq)^{-l}$ probability.
\end{theorem}

\subsection{Estimation of the Parameter $\theta$}\label{lowerbound}
To guarantee the solution returned by Algorithm \ref{alg3} is a $(1-1/e-\varepsilon)$-approximate solution for the PM-$\widetilde{G}$ problem with high probability, the number $\theta$ of the random RR sets generated in Algorithm \ref{alg1} should satisfy inequality (\ref{eqn3}).
For simplicity, we define
\begin{equation}\label{eqn12}
\setlength{\abovedisplayskip}{1pt}
\setlength{\belowdisplayskip}{1pt}
\lambda=(8q+2\varepsilon)nq^{2}p_{max}\cdot (l\log nq+\log (2qk^{*})+\log \tbinom{nq}{k^{*}})\cdot \varepsilon^{-2}
\end{equation}
and rewrite (\ref{eqn3}) as
\begin{equation}\label{eqn13}
\setlength{\abovedisplayskip}{1pt}
\setlength{\belowdisplayskip}{1pt}
\theta \ge \lambda/\textit{OPT}.
\end{equation}

However, since \textit{OPT} is unknown in advance, it is difficult to set $\theta$ directly based on (\ref{eqn13}). Inspired by the technique used in \cite{Tang14}, we address this challenge by finding an estimate $u$ of \textit{OPT} which is also a lower bound of \textit{OPT}. Then, by setting $\theta=\lambda/u$, we can guarantee $\theta$ satisfying inequality (\ref{eqn13}). On the other hand, $\theta$ should be set reasonably small in order to avoid time overheads, which requests the lower bound $u$ to be as close to \textit{OPT} as possible.

\subsubsection{First attempt}\label{first attempt}
In this section, an estimation of  \textit{OPT} is presented, which is based on the results in \cite{Tang14}. Though this estimation is not good enough, we remain it here in order to evaluate the time complexity of our algorithms.

Define the \emph{width} of an RR set $R$, denoted by $\omega(R)$, as the number of directed edges in $\widetilde{G}$ which point to the nodes in $R$. That is, $\omega(R)=\sum_{v\in R}(\text{the in-degree of}\ v \ \text{in}\ \widetilde{G})$. Obviously, if an edge is examined in the generation of $R$, then it must point to a node in $R$. Let \textit{EW} be the \emph{expected width} of a random RR set, that is, the expected number of coin tosses required to generate a random RR set. Therefore, it can be easy to verify that the expected time complexity of Algorithm \ref{alg1} is $O(\theta\cdot EW)$.

The connection between \textit{EW} and the expected spread of any node in $\mathcal{V}^{*}$ is formalized in the following lemma \cite{Tang14}.

\begin{lemma}\label{lem3}
$\frac{n}{m}\textit{EW}=\mathbb{E}[I(\{v^{*}\})]$, where the expectation of $I(\{v^{*}\})$ is taken over the randomness in $v^{*}$ and the influence propagation process.
\end{lemma}

Lemma \ref{lem3} implies that $p_{min}\cdot\frac{n}{m}\textit{EW}\le \textit{OPT}$, since $\mathbb{E}[I(S^{*})]$ is the expected spread of at least $\lfloor B/c_{max}\rfloor$ seed nodes and the profit of activating any node in $\mathbb{E}[I(S^{*})]$ is at least $p_{min}$. As $p_{min} \cdot \frac{n}{m}\textit{EW}$ is easy to be estimated, we can choose $u=p_{min} \cdot \frac{n}{m}\textit{EW}$ as a lower bound of $\textit{OPT}$. However, when $|S^{*}| \gg 1$, $u=p_{min} \cdot \frac{n}{m}\textit{EW}$ renders $\theta=\lambda /u$ unnecessarily large and makes $u=p_{min} \cdot \frac{n}{m}\textit{EW}$ an unfavorable choice of $u$. 

\subsubsection{Better lower bound of \textit{OPT}} \label{better bound}
Now we consider another closer estimation of \textit{OPT}. For $i= 1,2,\ldots,q$, we consider an extreme situation of the PM-$\widetilde{G}$ problem in which the seed set only contains the nodes in $G^{(i)}$. In such situation, the seed set consists of no more than $k_{i}=\lfloor B/c_{i}\rfloor$ nodes and the PM-$\widetilde{G}$ problem turns to seek for a size-$k_{i}$ seed set with the maximum profit, which is equivalent to the $k_{i}$-size IM problem in $G^{(i)}$. According to the results in \cite{Tang14}, the IM problem in $G$ under the IC model can be solved by an algorithm, called \textit{TIM}$^{+}$. 

The \emph{TIM$^{+}$} algorithm based on the RIS technique consists of two phases. The first phase, called parameter estimation, receives an estimate \emph{KPT$^{+}$} of the optimum and uses it to compute $\theta'$ which is the number of the random RR sets needed to generate. The second phase, called node selection, samples $\theta'$ random RR sets from $G$ and applies the greedy algorithm to derive a size-$k$ node set $S_{k}$ covering a large number of RR sets. (The details of the \textit{TIM}$^{+}$ algorithm can be seen in the appendix.)

We use the \textit{TIM}$^{+}$ algorithm to solve the $k_{i}$-size IM problem in $G^{(i)}$, and obtain an approximate solution (denoted by $S_{k_{i}}$) and the estimation of its expected spread (denoted by $\hat{\sigma}(S_{k_{i}})$). Then we compute the corresponding estimation of profit (denoted by $\overline{u}_{i}$) when $S_{k_{i}}$ is used as seed set. Based on the analysis in \cite{Tang14}, we compute ${u}_{i}=\overline{u}_{i}/(1+\frac{\epsilon'}{2})$ to ensure that ${u}_{i}\leq OPT$ with high probability. Then take $\max_{1\le i \le q}\{{u}_{i}\}$ as an estimation of \textit{OPT}, which is also a lower bound of \textit{OPT} with high probability. The estimating procedure is detailed in Algorithm \ref{alg5}, and the main result of \textit{TIM}$^{+}$ is presented as follows.
\vspace{-0.05in}
\begin{itemize}
\setlength{\itemsep}{0pt}
\setlength{\parsep}{0pt}
\setlength{\parskip}{0pt}
\item \textbf{Input :} A graph $G$, the constraint number $k$ of the seed set, a constant $l'>0$ and a parameter $\varepsilon'\in(0,1)$.
\item \textbf{Output :} A seed set $S_{k}$ and the estimation $\hat{\sigma}(S_{k})$ of its expected spread.
\item \textbf{Approximation}: $S_{k}$ is a $(1-1/e-\varepsilon')$-approximate solution of the IM problem, with at least $1-3n^{-l^{'}}$ probability.
\item \textbf{Time complexity :} $O\bigr((k+l')(m+n)\log n/(\varepsilon')^{2}\bigr)$.
\end{itemize}
\vspace{-0.05in}

Let $S_{k}^{*}$ be the optimal solution, and $\sigma(S_{k}^{*})$ be the optimum of the $k$-size IM problem in $G$. Based on Lemma $7$ and Lemma $8$ in \cite{Tang14}, we have:

\begin{lemma}\label{eqn10}
Let $S_{k}$ be the solution returned by the \textit{TIM}$^{+}$ algorithm for the $k$-size IM problem, and $\hat{\sigma}(S_{k})$ be the estimation of the expected spread of $S_{k}$, then
\[
\setlength{\abovedisplayskip}{2pt}
\setlength{\belowdisplayskip}{1pt}
\emph{Pr}\left[(1-1/e)(1-\varepsilon'/2)\sigma(S_{k}^{*})< \hat{\sigma}(S_{k})<(1+\varepsilon'/2)\sigma(S_{k}^{*})\right]
> 1-4n^{-l'}.
\]
\end{lemma}

\begin{algorithm}[t]
\caption{OPT Estimation} 
\label{alg5}
\hspace*{0.02in} {\bf Input:} 
Graph $\widetilde{G}$, a budget $B$, $c_{1},c_{2},\ldots,c_{q}$, $p_{1},p_{2},\ldots,p_{q}$ and $0< \varepsilon' < 1$, $l'>0$.\\
\hspace*{0.02in} {\bf Output:} 
A lower bound $u^{*}$ of \textit{OPT}.
\begin{algorithmic}[1]
\For{$i$ from $1$ to $q$}
¡¡¡¡\State $k_{i}=\lfloor \frac{B}{c_{i}}\rfloor$;¡¡¡¡
¡¡¡¡\State $\hat{\sigma}(S_{k_{i}})\leftarrow$ $\textit{TIM}^{+}(G^{(i)},k_{i}, \varepsilon',l')$;¡¡¡¡
¡¡¡¡\State ${u}_{i}=p_{i}\cdot \hat{\sigma}(S_{k_{i}})/(1+\varepsilon'/2)$;¡¡¡¡
\EndFor
\State \Return $u^{*}=\max_{1\le i \le q}\{u_{i}\}$.
\end{algorithmic}
\end{algorithm}

\begin{theorem}\label{thm2}
Algorithm \ref{alg5} returns
\[
\setlength{\abovedisplayskip}{1pt}
\setlength{\belowdisplayskip}{1pt}
u^{*}\in \left[\frac{(1-1/e)(1-\varepsilon'/2)}{(1+\varepsilon'/2)q}\textit{OPT},\textit{OPT}\right]
\]
with at least $1-4qn^{-l'}$ probability and runs in $O\bigr((k^{*}+l')(m+n)q\log n/(\varepsilon')^{2}\bigr)$ expected time, where $l'>0$ and $0<\varepsilon'<1$.
\end{theorem}

\subsubsection{Refined Estimation of \textit{OPT}}\label{Extensions}
In this section, we present another method to estimate \textit{OPT}. Clearly, the efficiency of the \emph{RMG} algorithm highly depends on the value of $u^{*}$ obtained by Algorithm \ref{alg5}. Though we could ensure that the output $u^{*}$ of Algorithm \ref{alg5} is no smaller than $\frac{(1-1/e)(1-\varepsilon'/2)}{(1+\varepsilon'/2)q}\cdot \textit{OPT}$ with high probability, $u^{*}$ may be much smaller than \textit{OPT} in experiments.

We pose an efficient solution (Algorithm \ref{alg6}) to the above problem, which adds an intermediate step between Algorithm \ref{alg5} and Algorithm \ref{alg3}. At first, we construct two matrices $P$ called profit matrix and $A$ called seed set matrix as follows (Algorithm \ref{alg6}, Lines 1-9). These two matrices both have $q$ rows and $k^{*}$ columns. For $i=1,2,\ldots,q$ and $j=1,2,\ldots,k_{i}$, the entry $a_{ij}$ of $A$ denotes the seed set which achieves the maximum profit of selecting $j$ seed nodes from $G^{(i)}$, and let $p_{ij}$ of $P$ be the estimation of profit obtained by using $a_{ij}$ as the seed set. For $i=1,2,\ldots,q$ and $j=k_{i} +1,\ldots, k^{*}$, we set $p_{ij}=0$ and $a_{ij}=\emptyset$. Then each time we select the seed set $a_{ij}$ with the maximum ratio of profit $p_{ij}$ to its activating cost $c_{i}\cdot j$ in the entire matrix, which means that we add the nodes set $a_{i^{*}j^{*}}$ whose profit satisfies
\[
\setlength{\abovedisplayskip}{1pt}
\setlength{\belowdisplayskip}{1pt}
p_{i^{*}j^{*}}=\arg\max_{\substack{1\le i \le q, 1 \le j \le k^{*}}} \left\{\frac{p_{ij}}{c_{i}\cdot j} \right\},
\]
to the current seed set, while still ensuring the activation cost of the update seed set no more than $B$. After that, set all the entries in row $i^{*}$ to $0$. Repeat the above process until the overall activation cost of the seed set is more than $B$. Denote $\hat{S}$ as the final seed set obtained, and $u^{**}=\hat{\rho}(\hat{S})$ (Lines 10-15). The final output of Algorithm \ref{alg6} is $u'=\max \{u^{*},u^{**}\}$, a new lower bound of \textit{OPT} (Line 16).

\begin{algorithm}[t]
\caption{Refine OPT Estimation} 
\label{alg6}
\hspace*{0.02in} {\bf Input:} 
Graph $\widetilde{G}$, a budget $B$, $c_{1},c_{2},\ldots,c_{q}$, $p_{1},p_{2},\ldots,p_{q}$, $l'>0$ and $0< \varepsilon' < 1$.\\
\hspace*{0.02in} {\bf Output:} 
A lower bound $u'$ of \textit{OPT}.
\begin{algorithmic}[1]
\State Initialize two matrices $P$ and $A$;
\For{$i$ from $1$ to $q$}
    \For{$j$ from $1$ to $k_{i}$}
¡¡¡¡\State $\hat{\sigma}(S_{j}^{(i)})\leftarrow$ $\textit{TIM}^{+}(G^{(i)},j,\varepsilon',l')$;¡¡¡¡
¡¡¡¡\State $a_{ij}=S_{j}^{(i)}$;¡¡¡¡
¡¡¡¡\State $p_{ij}=p_{i}\cdot \hat{\sigma}(S_{j}^{(i)})$;¡¡¡¡
     \EndFor
     \For{$j$ from $k_{i}+1$ to $k^{*}$}
¡¡¡¡\State $a_{ij}=\emptyset$;¡¡¡¡
¡¡¡¡\State $p_{ij}=0$;¡¡¡¡
     \EndFor
\EndFor    
\State $\hat{S}\leftarrow \emptyset$;
\While{$c(\hat{S})\le B$}
\State $p_{i^{*}j^{*}}=\arg \max_{1\le i \le q,1\le j\le k^{*}}\, \{\frac{p_{ij}}{c_{i}\cdot j}\}$;
\State $\hat{S}=\hat{S}\cup a_{i^{*}j^{*}}$;
\State Set entries of row $i^{*}$ in matrix $P$ to $0$;
\EndWhile
\State  $u^{**}=\hat{\rho}(\hat{S})$;   
\State \Return $u'=\max \{u^{*},u^{**}\}$.
\end{algorithmic}
\end{algorithm}

\subsection{Putting It All Together}
In summary, our \emph{RMG} algorithm for the PM$^{2}$A problem works as follows.
Given the social network $G$, $B$, $c_{1},c_{2},\ldots,c_{q}$, $p_{1},p_{2},\ldots,p_{q}$, parameters $\varepsilon,\varepsilon'$, $l$ and $l'$, we first construct the graph $\widetilde{G}=G^{(1)}\cup G^{(2)}\cup \ldots \cup G^{(q)}$. Then \emph{RMG} implements Algorithm \ref{alg5} and obtains a value of $u^{*}$ in return. And then \emph{RMG} computes $\theta=\lambda/ u^{*}$ in which $\lambda$ is defined in (\ref{eqn12}) and invokes Algorithm \ref{alg1} to generate a set $\mathcal{R}$ of random RR sets. Finally, we run Algorithm 3 with $\widetilde{G}$, $\varepsilon$, $B$ and $\theta$ as the input and take its output $S_{\mathcal{A}}$ as the final result of the PM$^{2}$A problem.

In the rest of this section, we discuss the time complexity of \emph{RMG} algorithm. Based on previous discussions, the expected time complexity of Algorithm \ref{alg1} is $O(\theta\cdot EW)$. In \ref{first attempt}, we have obtained that $u=p_{min}\cdot \frac{n}{m}EW$ is a lower bound of \textit{OPT}. By setting $\theta=\lambda/u$, we can obtain that Algorithm \ref{alg1} has an expected time complexity of
\[ 
\setlength{\abovedisplayskip}{1pt}
\setlength{\belowdisplayskip}{1pt}
O(\theta \cdot \textit{EW})=O(\frac{m\lambda}{np_{min}})
=O\bigr((k^{*}+l+1)(m+n)q^{3}p_{max}\log (nq)/(p_{min}\cdot\varepsilon^{2})\bigr).
\]
Clearly, Algorithm \ref{alg2} runs in $O(q\theta)=O\bigr((k^{*}+l+1)(m+n)q^{4}p_{max}\log (nq)/(p_{min}\cdot\varepsilon^{2})\bigr)$ expected time.

Now we are in the position to analyse the expected running time of Algorithm \ref{alg3}. For any $S\in \Omega$, $\hat{\rho}(S)$ is computed by Algorithm \ref{alg2} which runs in $O\bigr((k^{*}+l+1)(m+n)q^{4}p_{max}\log (nq)/(p_{min}\cdot\varepsilon^{2})\bigr)$ expected time. The first part of Algorithm \ref{alg3} from line $2$ to $6$ invokes Algorithm \ref{alg2} at most  $(nq)^{2}$ times. The second part of Algorithm \ref{alg3} from line $7$ to $22$ invokes Algorithm \ref{alg2} at most  $k^{*}\cdot(nq)^{4}$ times. Thus, Algorithm \ref{alg3} has an expected time complexity of $O\bigr(k^{*}(k^{*}+l+1)(m+n)n^{4}q^{8}p_{max}\log (nq)/(p_{min}\cdot \varepsilon^{2})\bigr)$.

By Theorems \ref{thm1} and \ref{thm2}, \emph{RMG} runs in $O\bigr(k^{*}(k^{*}+l+1)(m+n)n^{4}q^{8}p_{max}\cdot \log (nq)/(p_{min}\cdot \varepsilon^{2})\bigr)$ expected time and returns a $(1-1/e-\varepsilon)$-approximate solution with at least $1-(nq)^{-l}-3qn^{-l'}$ probability.

\section{Experimental Evaluation}
\label{Experiments}
In this section, we show the effectiveness of our proposed algorithm on three social network datasets. The goal of the experiments is multifold. First, we would like to evaluate the performance of the \emph{RMG} algorithm as measured by the achieved expected total profit. Second, we evaluate the extent to which the estimated \textit{OPT} and refined \textit{OPT} estimate the lower bound of the profit, which indirectly control the efficiency of the profit maximization algorithm. Finally, we show the distribution of budget and profit produced by our algorithm for different products on multiple datasets to reveal the superiority of our algorithms in depth.

\subsection{Experimental Setup}

\begin{table}[hptb]\centering\small
\setlength{\abovecaptionskip}{-0.4cm}
\setlength{\belowcaptionskip}{-0cm}
\caption{Dataset characteristics}
\begin{tabular}{{|c|c|c|c|c|}}
\hline
\textbf{Dataset} & \textbf{\emph{n}} & \textbf{\emph{m}} & \textbf{Type} & \textbf{Average degree}\\
\hline
\emph{NetHEPT} & 15,229 & 31,376 & undirected & 4.1\\
\hline
\emph{wikiVote} & 7,115 & 103,689 & directed & 29.1\\
\hline
\emph{Epinions} & 75,879 & 508,837 & directed & 13.4\\
\hline
\end{tabular}
\label{tab:dataset}
\end{table}
\vspace{-0.15in}

\textbf{Datasets.} We conduct extensive experiments on three real benchmark social networks: \emph{NetHEPT}, \emph{wikiVote} and \emph{Epinions} to examine the effectiveness of the \emph{RMG} algorithm. Basic statistics of the datasets are summarized in Table \ref{tab:dataset}, where $n$ denotes the number of nodes and $m$ denotes the number of edges in the social graph. For undirected graphs, we reverse every edge in both directions so as to make each undirected edge into two directed edges. Note that the number of edges are doubled in this case. All datasets used in our experiments are publicly available at \cite{Leskovec2014}.

\textbf{Influence Model.} In this work, we adopt the standard Independent Cascade (IC) model as the influence model, which is widely used in the literature \cite{Tang14, Tang15}. As for the IC model, we set the propagation probability of each directed edge as reciprocal of the in-degree of the node that the edge points to. Specifically, for each edge $e$ we first identify the node $v$ that $e$ points to, and then set $p(e)=1/d(v)$, where $d(v)$ denotes the in-degree of $v$. This setting of $p(e)$ is widely used in prior works \cite{Tang14, Goyal, Jung}.

\textbf{Algorithms.} In addition to our proposed algorithm, we use three algorithms as baseline algorithms for comparison purpose, namely, \emph{Random}, \emph{Greedy}, and \emph{PMCE} \cite{Zhang16}. In particular, \emph{Random} is a baseline algorithm that randomly select nodes from the network and assign random product to each node while satisfying the budget constraint. \emph{Greedy} is an iterative procedure, the intuition behind is to select the pair of node and product with maximum ratio of the marginal increase in expected profit over the cost in each round, until the budget is exhausted. \emph{PMCE} is another baseline algorithm, i.e., the \emph{Profit Maximization with Cost Effectiveness} algorithm proposed in \cite{Zhang16}. It first constructs two candidate solutions and then select the better one as the final result. The first candidate is selected in an iterative greedy process. In each round, the node with maximum ratio of the marginal profit increase over the square of cost is selected, and the process runs until the budget is used up. It then finds the second candidate using a similar iterative greedy process, only with a different guideline to select a node in each iteration: selecting the node with maximum marginal profit increase. The intuition behind \emph{PMCE} is to consider both the cases that to emphasize the importance of product cost (first candidate) and to ignore the importance of product cost (second candidate). Note that \emph{PMCE} is designed under the Linear Threshold (LT) model only, but we incorporate the triggering model generalization technique \cite{Tang14} into \emph{PMCE} and extend it to the IC model.
\vspace{-0.2in}
\begin{table}[htbp]\centering\small
\caption{Product Statistics}
\begin{tabular}{{|c|c|l|l|l|}}
\hline
\textbf{Dataset}& \textbf{Product} & \textbf{Profit} & \textbf{Cost} & \textbf{Ratio}\\
\hline
\emph{NetHEPT}& P1;P2;P3 & 0.39;0.55;0.67 & 0.36;0.48;0.65 & 1.08;1.15;1.03\\
\hline
\emph{wikiVote} & P1;P2;P3 & 0.45;0.65;0.41 & 0.12;0.20;0.80 & 3.75;3.25;0.51\\
\hline
\emph{Epinions} &P1;P2;P3 & 0.45;0.20;0.06 & 0.08;0.65;0.78 & 5.63;0.31;0.08\\
\hline
\end{tabular}
\label{tab:prod}
\end{table}
\vspace{-0.15in}

\textbf{Parameters.} Unless otherwise specified, we set $\varepsilon=\varepsilon'=\bar{\varepsilon}=0.1$ ($\bar{\varepsilon}$ is another error parameter in the \emph{TIM}$^{+}$ alg.) and $B=15$. For our solutions, we set $l$ and $l'$ in a way that ensures a success probability of $1-1/n$. For the baseline algorithms, we set the number of Monte Carlo simulations to $r=10^4$, following the standard practice in the literature \cite{Tang14}. In our experiments, we randomly generate  the profit and cost of the products for each dataset in three cases. First, the profit over cost ratio is similar among different products (\emph{NetHEPT}). Second, we set two products with higher profit over cost ratio than the other product (\emph{wikiVote}). Third, we set a single product with higher cost than the other two products (\emph{Epinions}). The product statistics are presented in Table \ref{tab:prod}. In later paragraphs, we provide a detailed analysis of different budget distribution patterns shown in these cases and make a deep exploration of the superiority of our proposed algorithms.

All experiments were run on a machine with Intel Xeon 2.40GHz CPU and 64GB memory, running 64-bit RedHat Linux server. For each set of experiments, we run the simulation for 100 rounds and average results are reported as follow.

\subsection{Experimental Results}

\begin{figure}[hptb]\centering
\vspace{-0.3cm}  
\setlength{\abovecaptionskip}{-0.1cm}   
\setlength{\belowcaptionskip}{-0.3cm} 
\includegraphics[scale=0.18]{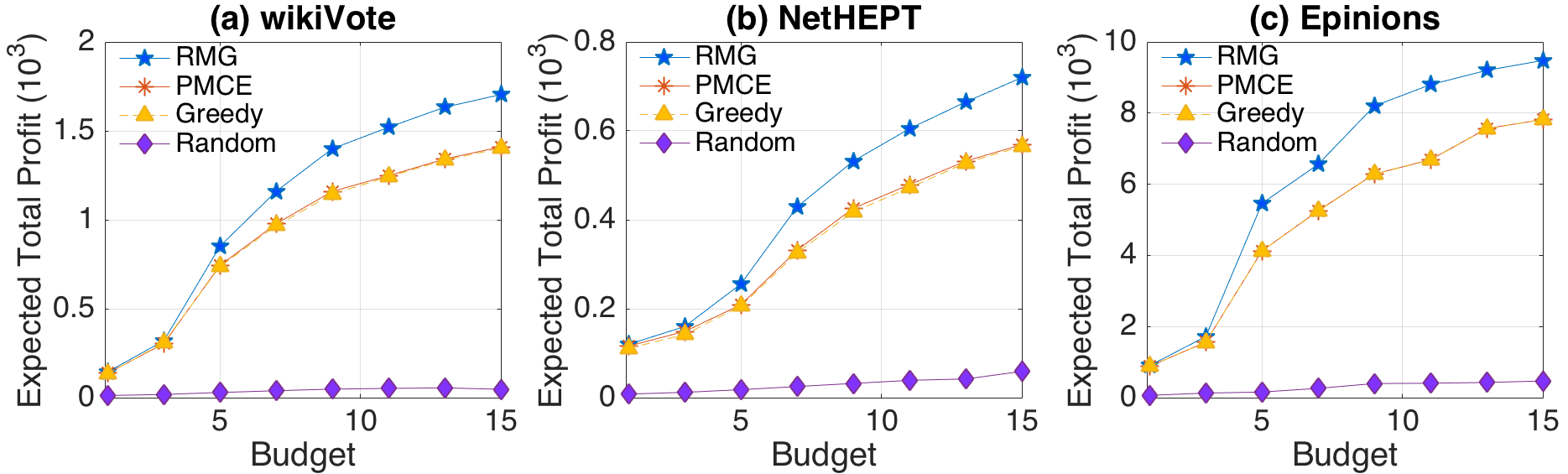}
\caption{Expected total profit vs. amount of budget.}\centering
\label{fig:profit}
\end{figure}

\textbf{Expected Total Profit.} Our first set of experiments compares our solutions in terms of expected total profit with baseline algorithms \emph{Random}, \emph{Greedy} and \emph{PMCE}. Figure \ref{fig:profit} shows the expected total profit yielded by each method on all tested datasets, with $B$ varying from $1$ to $15$. The $x$-axis holds the amount of budget and the $y$-axis holds the expected total profit. We observe that the trend of \emph{PMCE} is almost in line with the trend of Greedy, and $RMG$ consistently outperforms all baseline algorithms. In particular, when $B=15$, \emph{RMG} leads \emph{PMCE} by over $20\%$ gain on all datasets, and the gap between \emph{RMG} and baseline algorithms becomes larger as the budget increases.

\begin{figure}[hptb]\centering
\vspace{-0.3cm}  
\setlength{\abovecaptionskip}{-0.1cm}   
\setlength{\belowcaptionskip}{-0.3cm} 
\includegraphics[scale=0.24]{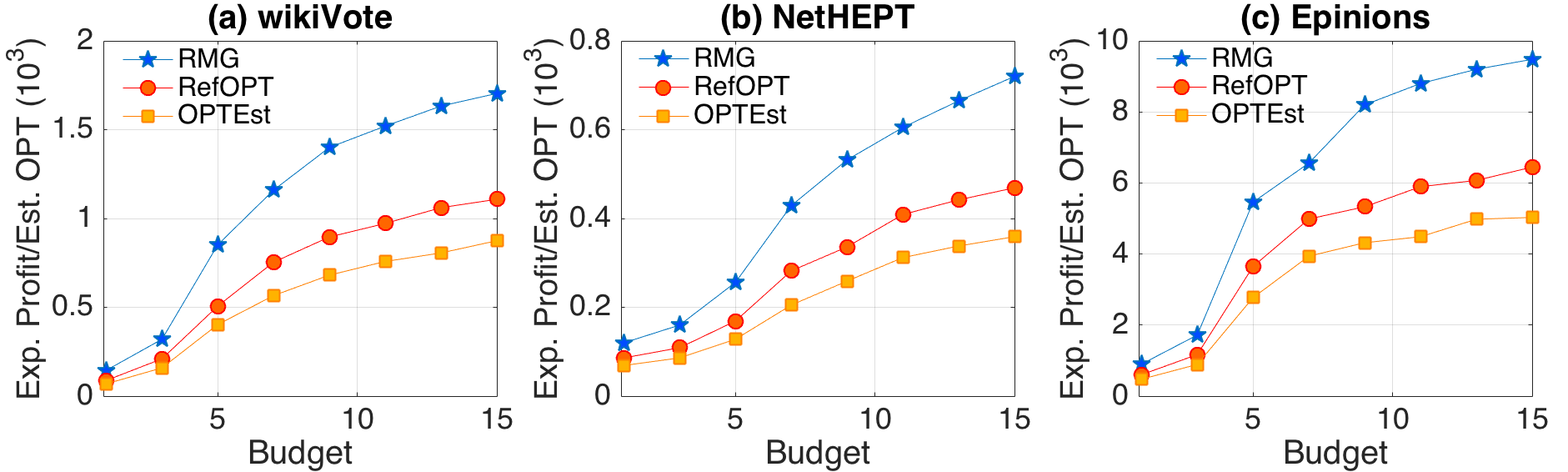}
\caption{Comparison of total profit with estimated OPT (Algorithm \ref{alg5}) and refined OPT (Algorithm \ref{alg6}).}\centering
\label{fig:opt}
\end{figure}

\vspace{-0.1in}
\textbf{Estimation on \textit{OPT}.} Figure \ref{fig:opt} presents the comparison of the expected total profit yielded by \emph{RMG} with the estimated OPT yielded by Alg. \ref{alg5} (OPTEst) and Alg. \ref{alg6} (RefOPT) respectively. The $x$-axis holds the amount of budget. For \emph{RMG}, the $y$-axis holds the expected total profit; and for OPTEst and RefOPT, the $y$-axis holds the estimated lower bound of \textit{OPT}. The budget $B$ ranges from $1$ to $15$. We observe that RefOPT produces a tighter estimation of the lower bound of \emph{OPT} on all datasets over a varying budget. This indicates that it is beneficial to incorporate the computation of maximum profit that can be achieved considering all possible combinations of budget distributions over multiple products. Thus Algorithm \ref{alg6} provides a sophisticated yet effective estimation on the lower bound of \textit{OPT}, leading to a higher efficiency of \emph{RMG}.

\begin{figure}[htbp]\centering
\vspace{-0.3cm}  
\setlength{\abovecaptionskip}{-0.0cm}   
\setlength{\belowcaptionskip}{-0.4cm} 
\includegraphics[scale=0.18]{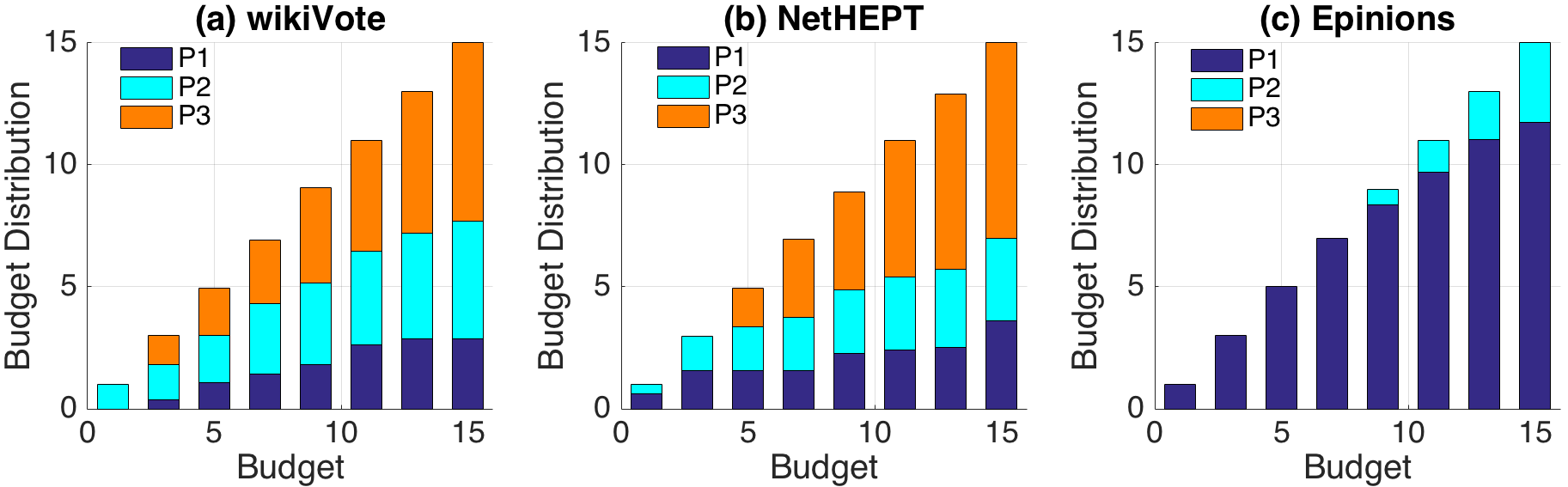}\\
\vspace{0.05in}
\includegraphics[scale=0.18]{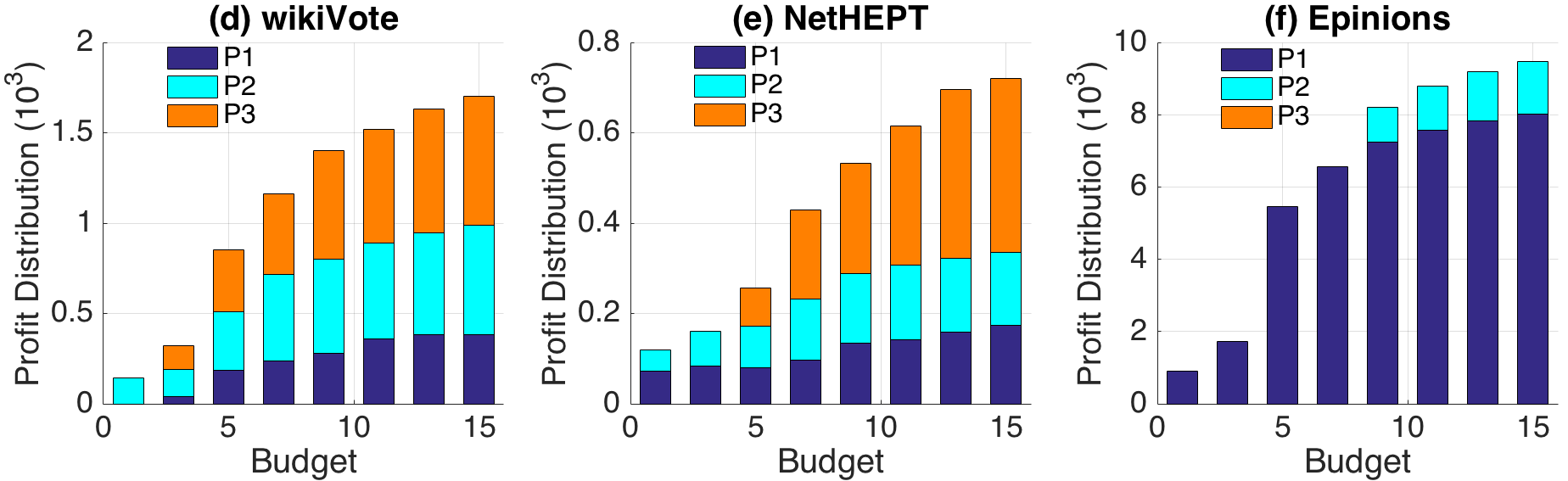}
\vspace{-0.05in}
\caption{Budget \& profit distributions.}\centering
\label{fig:dist}
\end{figure}

\textbf{Budget \& Profit Distribution.} We take a further step to explore the distribution of budget and profit produced by \emph{RMG} with a varying budget. Figure \ref{fig:dist} illustrates how the budget is distributed over multiple products and the corresponding profit gained from each product with the budget varying from $1$ to $15$. We observe that with a limited budget at the very beginning, all the budget is spent on promoting the product with highest profit cost ratio. As the budget increases, spending more on a single profitable product is not preferred and gradually adjusting budget distribution over multiple products becomes crucial. Thus \emph{RMG} balances the cost and profit in a long run and produces a distribution that maximizes the profit.

In summary, our experiments on various settings demonstrate that the \emph{RMG} algorithm is effective, producing far superior solutions than the baselines.

\section{Conclusion}
\label{Conclusion}
Traditional Influence maximization problem focuses on the diffusion of a single product or information in the social network, aiming to seek for a small node set of maximum influence. However, in reality, one company may produce several products to meet  the demand of customers. The PM$^{2}$A problem considers the diffusion of multiple different products in the social network, and seeks for a seed set within the limited budget to achieve the goal of profit maximization. Therefore, how to allocate the limited budget among multiple  products is crucial for the company in designing commercial activities.

In this paper, we propose a \emph{RMG} algorithm for the PM$^{2}$A problem. The algorithm runs in $O(k^{*}(k^{*}+l+1)(m+n)n^{4}q^{8}p_{max}\log (nq)/(p_{min}\cdot \varepsilon^{2}))$ expected time and returns a $(1-1/e-\varepsilon)$-approximate solution with at least $1-(nq)^{-l}-3qn^{-l'}$ probability, which significantly  improves upon prior works in terms of performance guarantee and is also the best performance ratio of the PM$^{2}$A problem even for one product. Experimental results on real-world social networks show that our \emph{RMG} algorithm outperforms the algorithm proposed in \cite{Zhang16} and other heuristics in terms of profit maximization, and could better allocate the budget. For future work, we plan to improve the \emph{RMG} algorithm in terms of the time complexity, and investigate the case in which multiple products spread in the social network and could compete with each other.

\bigskip

\noindent{\bf Acknowledgement.}  This work was supported in part by National Natural Science Foundation of China (11501316), China Postdoctoral Science Foundation (2016M600556), Qingdao Postdoctoral Application Research Project (2016156), and Natural Science Foundation of Shandong Province of China (ZR2017QA010).

\appendix
\section{Proof of some conclusions}

\noindent
\textbf{Proof of Lemma \ref{lem2}.}
For any seed set $S\in\Omega$, let $\mu_{i}=\sigma(S^{(i)})/(nq)=\mathbb{E}[F_{\mathcal{R}}(S^{(i)})]$, which represents the probability that $S^{(i)}$ overlaps with a random RR set.

Then $\theta \cdot F_{\mathcal{R}}(S^{(i)})$ can be regarded as the sum of $\theta$ i.i.d. Bernoulli variables with a mean $\mu_{i}$. Thus, we have
\begin{flalign}
&\text{Pr}\left[|\hat{\rho}(S^{(i)})-\rho(S^{(i)})|\ge \frac{\varepsilon}{2q}\cdot \textit{OPT} \right]\nonumber\\
=&\text{Pr}\left[|\theta \cdot F_{\mathcal{R}}(S^{(i)})-\theta \cdot \mu_{i}|\ge \frac{\varepsilon \cdot \textit{OPT}}{2nq^{2}p_{i}\mu_{i}}\cdot \theta \mu_{i}\right]\label{eqn5}.
\end{flalign}

Let $\delta=\frac{\displaystyle\varepsilon \cdot \textit{OPT}}{\displaystyle2nq^{2}p_{i}\mu_{i}}$. By Chernoff bounds, inequality (\ref{eqn3}) and the fact that $\rho(S^{(i)})=p_{i}\sigma(S^{(i)})=p_{i}\cdot nq \mu_{i}\le \textit{OPT}$, the following inequality holds for the right hand side (r.h.s.) of (\ref{eqn5}): 
 \begin{flalign}
\text{r.h.s. of (\ref{eqn5})}&<2\exp \{-\frac{\delta^{2}}{2+\delta}\cdot \theta \mu_{i}\}\nonumber\\
                      &=2\exp \{-\frac{\varepsilon^{2}\cdot \textit{OPT}^{2}}{2nq^{2}p_{i}(4nq^{2}p_{i}\mu_{i}+\varepsilon\cdot \textit{OPT})}\cdot \theta\}\nonumber\\
                      &\le 2\exp \{-\frac{\varepsilon^{2}\cdot \textit{OPT}^{2}}{2nq^{2}p_{i}(4q\cdot \textit{OPT}+\varepsilon\cdot \textit{OPT})}\cdot \theta\}\nonumber\\
                      &\le (nq)^{-l}(qk^{*})^{-1}/ \binom{nq}{k^{*}}\label{eqn14}.
 \end{flalign}
Furthermore, we have
 \begin{flalign}
     &\text{Pr}\left[\left|\hat{\rho}(S)-\rho(S)\right|< \frac{\varepsilon}{2}\cdot \textit{OPT}\right]\nonumber\\
    =&\text{Pr}\left[\left|\sum_{i=1}^{q}(\hat{\rho}(S^{(i)})-\rho(S^{(i)}))\right|< \frac{\varepsilon}{2}\cdot \textit{OPT}\right]\nonumber\\
  \ge& \text{Pr}\left[\sum_{i=1}^{q}\left|\hat{\rho}(S^{(i)})-\rho(S^{(i)})\right|<\frac{\varepsilon}{2}\cdot \textit{OPT}\right]\nonumber\\
  \ge& \text{Pr}\left[q\cdot \max_{1\le i\le q}|\hat{\rho}(S^{(i)})-\rho(S^{(i)})|<\frac{\varepsilon}{2}\cdot \textit{OPT}\right]\nonumber\\
    =& \text{Pr}\left[|\hat{\rho}(S^{(1)})-\rho(S^{(1)})|<\frac{\varepsilon}{2q}\cdot \textit{OPT},\ldots,
     |\hat{\rho}(S^{(q)})-\rho(S^{(q)})|<\frac{\varepsilon}{2q}\cdot \textit{OPT}\right]\label{eqn15}.
 \end{flalign}
 By the union bound and Equation (\ref{eqn14}), we have
 \[
 \begin{aligned}
\text{r.h.s. of (\ref{eqn15})} \ge& \sum_{i=1}^{q}{\text{Pr}\left[|\hat{\rho}(S^{(i)})-\rho(S^{(i)})|<\frac{\varepsilon}{2q}\cdot \textit{OPT}\right]}-(q-1)\\
  \ge& 1-(nq)^{-l}(k^{*})^{-1}/\binom{nq}{k^{*}}
  \end{aligned}
  \]
Therefore, the lemma is proved. $\qed$\\

\noindent
\textbf{Proof of Lemma \ref{lem4}.} For any node set $S\subseteq \widetilde{V}$, adding nodes from $\widetilde{V}\setminus S$ into $S$ can never decrease $F_{\mathcal{R}}(S^{(i)})$ ($i=1,2,\ldots,q$). Hence, $\hat{\rho}(S)=nq\sum_{i=1}^{q}{p_{i}F_{\mathcal{R}}(S^{(i)})}$ is nondecreasing since $p_{i}>0$.

For any $S\subseteq T\subseteq \widetilde{V}$ and $y\in \widetilde{V}\setminus T$, we will prove the following inequality (\ref{eqn8}) which implies $F_{\mathcal{R}}(\cdot)$ is submodular:

\begin{equation}\label{eqn8}
   F_{\mathcal{R}}(S\cup \{y\})-F_{\mathcal{R}}(S)\ge F_{\mathcal{R}}(T\cup \{y\})-F_{\mathcal{R}}(T).
\end{equation}

Let $W_{1}$, $W_{2}$, $W_{3}$ and $W_{4}$ be the sets of RR sets in $\mathcal{R}$ covered by $T\cup \{y\}$, $T$, $S\cup \{y\}$ and $S$, respectively. Then $W_{1}\backslash W_{2}$ represents the set of RR sets which can be covered by $\{y\}$ but not covered by $T$, and $W_{3}\backslash W_{4}$ represents the set of RR sets which can be covered by $\{y\}$ but not covered by $S$. Recall that $S\subseteq T$, we have $(W_{1}\backslash W_{2})\subseteq (W_{3}\backslash W_{4})$.
By the definition of $F_{\mathcal{R}}(\cdot)$, $F_{\mathcal{R}}(S\cup \{y\})-F_{\mathcal{R}}(S)$ represents the proportion of RR sets in $\mathcal{R}$ which can be covered by $\{y\}$ but not covered by $S$. It follows that inequality (\ref{eqn8}) holds, according to the relationship between $(W_{1}\backslash W_{2})$ and $(W_{3}\backslash W_{4})$.
Therefore, $\hat{\rho}(S)=nq\sum_{i=1}^{q}{p_{i}F_{\mathcal{R}}(S^{(i)})}$ is submodular. $\qed$\\

\noindent
\textbf{Proof of Theorem \ref{thm1}.} Let $S_{\mathcal{A}}$ be the node set returned by Algorithm \ref{alg3}, and $S_{p}^{*}$ be the optimal solution of problem (\ref{eqn7}). As $S_{\mathcal{A}}$ is obtained by a $(1-1/e)$-approximation algorithm for problem (\ref{eqn7}) \cite{Leskovec}, we have $\hat{\rho}(S_{\mathcal{A}})\ge (1-1/e)\hat{\rho}(S_{p}^{*})$. Recall that $S^{*}$ is the optimal solution for the PM-$\widetilde{G}$ problem and $\textit{OPT}=\rho(S^{*})$, we have $\hat{\rho}(S_{p}^{*})\ge \hat{\rho}(S^{*})$, leading to $\hat{\rho}(S_{\mathcal{A}})\ge (1-1/e)\hat{\rho}(S^{*})$.

According to Lemma \ref{lem2}, inequality (\ref{eqn4}) holds with at least $1-(nq)^{-l}(k^{*})^{-1}/ \binom{nq}{k^{*}}$ probability for a given seed set $S\in \Omega$. By the assumption that $k^{*}\le \lfloor nq/2\rfloor$, we can obtain that $|\Omega|\le \sum_{1\le j\le k^{*}}{\binom{nq}{j}}\le k^{*}\cdot \binom{nq}{k^{*}}$. Then, by the union bound, inequality (\ref{eqn4}) holds simultaneously for all node sets belonging to $\Omega$ with at least $1-(nq)^{-l}$ probability. In that case, we have
\[
 \begin{aligned}
  \rho(S_{\mathcal{A}})&> \hat{\rho}(S_{\mathcal{A}})-\frac{\varepsilon}{2} \cdot OPT \ge (1-\frac{1}{e})\hat{\rho}(S^{*})-\frac{\varepsilon}{2} \cdot OPT\\
                           &> (1-\frac{1}{e})(1-\frac{\varepsilon}{2})\cdot OPT-\frac{\varepsilon}{2} \cdot OPT>(1-\frac{1}{e}-\varepsilon)\cdot OPT.
 \end{aligned}
\]
Thus, Theorem \ref{thm1} is proved. $\qed$\\

\noindent
\textbf{Proof of Lemma \ref{eqn10}.} According to Lemma $7$ and Lemma $8$ in \cite{Tang14}, we have that 
\[\text{Pr}\left[\emph{KPT}^{*}\in \left[\frac{\emph{KPT}}{4}, \sigma(S_{k}^{*})  \right] \right]\ge 1-n^{-l'}, \] 
and
\[\text{Pr}\left[\emph{KPT}^{+}\in \left[\emph{KPT}^{*}, \sigma \left(S_{k}^{*}\right)\right] \mid \emph{KPT}^{*}\in \left[\frac{\emph{KPT}}{4}, \sigma(S_{k}^{*}) \right] \right]\ge 1-n^{-l'}.\]
 Thus,
\[\text{Pr}\left[\emph{KPT}^{+}\le \sigma(S_{k}^{*})\right]\ge \text{Pr}\left[\emph{KPT}^{+}\in \left[\emph{KPT}^{*}, \sigma \left(S_{k}^{*}\right)  \right] \right]\ge 1-2n^{-l'}. \]

Let $\lambda'=(8+2\varepsilon')n(l'\log n+\log 2+\log \binom{n}{k})\cdot (\varepsilon')^{-2}$ and $\theta'=\lambda'/KPT^{+}$, then $\text{Pr}\left[\theta'\ge \lambda'/\sigma(S_{k}^{*})\right]\ge 1-2n^{-l'}$.

For any size-$k$ node set $S_{k}$, suppose that $\theta'$ satisfies $\theta'\ge \lambda'/\sigma(S_{k}^{*})$, then $|\hat{\sigma}(S_{k})-\sigma(S_{k})|<(\varepsilon'/2)\cdot \sigma(S_{k}^{*})$ holds with at least $1-n^{-l'}/\binom{n}{k}$ probability.

It follows that when $\theta'$ satisfies $\theta'\ge \lambda'/\sigma(S_{k}^{*})$, \[\text{Pr}\left[\hat{\sigma}(S_{k})<(1+\frac{\varepsilon'}{2})\sigma(S_{k}^{*})\right]\ge 1-n^{-l'}/\binom{n}{k}\]
 and
 \[\text{Pr}\left[\hat{\sigma}(S_{k})> (1-\frac{1}{e})(1-\frac{\varepsilon'}{2})\sigma(S_{k}^{*})\right]\ge 1-n^{-l'}/\binom{n}{k}.\] Thus,
\[
\begin{aligned}
&\text{Pr}\left[ (1-\frac{1}{e})(1-\frac{\varepsilon'}{2})\sigma(S_{k}^{*})< \hat{\sigma}(S_{k})<(1+\frac{\varepsilon'}{2})\sigma(S_{k}^{*})\mid \theta'\ge \frac{\lambda'}{\sigma(S_{k}^{*})}\right]
\ge 1-2n^{-l'}/\tbinom{n}{k}.
\end{aligned}
\]
Therefore, we have
\[
\begin{aligned}
&\text{Pr}\left[(1-1/e)(1-\varepsilon'/2)\sigma(S_{k}^{*})< \hat{\sigma}(S_{k})<(1+\varepsilon'/2)\sigma(S_{k}^{*})\right]\\
&\ge (1-2n^{-l'})(1-2n^{-l'}/\tbinom{n}{k})> 1-4n^{-l'}.
\end{aligned}
\] $\qed$\\

\noindent
\textbf{Proof of Theorem \ref{thm2}.} Recall that the optimal solution of the PM-$\widetilde{G}$ problem is $S^{*}$, denote $S^{*}=\bar{S}^{(1)}\cup \bar{S}^{(2)}\cup \cdots \cup \bar{S}^{(q)}$. Then, we obtain that 
\[\textit{OPT}=\rho(S^{*})=\sum_{i=1}^{q}{\rho(\bar{S}^{(i)}})=\sum_{i=1}^{q}{p_{i}\cdot \sigma(\bar{S}^{(i)})}.\]
By the definition of the optimal solution, we have $\bar{S}^{(i)}$ is the optimal solution of the $|\bar{S}^{(i)}|$-size IM problem in $G^{(i)}$. For the $k_{i}$-size IM problem in $G^{(i)}$ ($i=1,2,\ldots,q$), let $S_{k_i}$ and $\hat{\sigma}(S_{k_i})$ be the
$(1-1/e-\varepsilon')$-approximate solution and the corresponding value returned by \textit{TIM}$^{+}$, respectively. Let $S_{k_{i}}^{*}$ be the optimal solution, and $\sigma(S_{k_{i}}^{*})$ be its expected spread. Based on Lemma \ref{eqn10}, we have\\
 \[
 \text{Pr}\left[(1-1/e)(1-\varepsilon'/2)\sigma(S_{k_{i}}^{*})< \hat{\sigma}(S_{k_{i}})<(1+\varepsilon'/2)\sigma(S_{k_{i}}^{*})\right]>1-4n^{-l'}.
 \]

It follows directly that $|S_{k_{i}}^{*}|\ge |\bar{S}^{(i)}|$ by the definitions of $k_{i}$ and $\bar{S}^{(i)}$. Recall that $S_{k_{i}}^{*}$ and $\bar{S}^{(i)}$ are the optimal solutions of the $k_i$-size and $|\bar{S}^{(i)}|$-size influence maximization problem in $G^{(i)}$, respectively. Then we have $\sigma(S_{k_{i}}^{*})\ge \sigma(\bar{S}^{(i)})$ since $\sigma(\cdot)$ is nondecreasing.

Let $u_{i}=p_{i}\cdot \hat{\sigma}(S_{k_{i}})/(1+\varepsilon'/2)$ and $u^{*}=\max_{1 \le i \le q}\{u_{i}\}$, then
 \begin{equation}\label{eqn9}
 \text{Pr}\left[\frac{(1-1/e)(1-\varepsilon'/2)}{1+\varepsilon'/2}\rho(S_{k_{i}}^{*})< u_{i}\le\textit{OPT} \right]> 1-4n^{-l'}.
 \end{equation}
Therefore, (\ref{eqn9}) holds simultaneously for all $i=1,2,\ldots,q$ with at least $1-4qn^{-l'}$ probability, which means that
\[
\textit{OPT}\le \sum_{i=1}^{q}{p_{i}\cdot \sigma(S_{k_{i}}^{*})}=\sum_{i=1}^{q}\rho(S_{k_{i}}^{*})< \frac{(1+\varepsilon'/2)q}{(1-1/e)(1-\varepsilon'/2)}u^{*},
\]
and $u^{*}\le\textit{OPT}$ hold simultaneously with at least $1-4qn^{-l'}$ probability.
In conclusion, $u^{*}\in \left[\frac{(1-1/e)(1-\varepsilon'/2)}{(1+\varepsilon'/2)q}\textit{OPT},\textit{OPT}\right]$ with at least $1-4qn^{-l'}$ probability.

\textit{TIM}$^{+}$ runs in $O\bigr((k_{i}+l')(m+n)\log n/(\varepsilon')^{2}\bigr)$ expected time, where $k_{i}=\lfloor B/c_{i} \rfloor$ is the number budget of the seed set \cite{Tang14}. Therefore, Algorithm \ref{alg5} runs in $O\bigr((k^{*}+l')(m+n)q\log n/(\varepsilon')^{2}\bigr)$ expected time. $\qed$\\

\section{Brief Introduction of the \emph{TIM$^{+}$} Algorithm}\label{appendix}
In this section, we give an outline of the \emph{TIM$^{+}$} algorithm. The \emph{TIM$^{+}$} algorithm based on the RIS technique consists of two phases. The first phase called parameter estimation receives an estimate \emph{KPT$^{+}$} of the optimum and uses it to compute $\theta^{'}$ which is the number of RR sets needed to generate. The second phase, called node selection, samples $\theta^{'}$ RR sets from $G$ and applies the greedy algorithm to derive a size-$k$ node set $S_{k}$ covering a large number of RR sets. We put all the algorithms in the entire process together in algorithm \ref{alg4}.


\begin{algorithm}[t]
\caption{$TIM^{+}$} 
\label{alg4}
\hspace*{0.02in} {\bf Input:} 
Graph $G$, a positive integer $k$ and $0<\bar{\varepsilon},\varepsilon'<1$.\\
\hspace*{0.02in} {\bf Output:} 
A value $\hat{\sigma}(S_{k})$ where $S_{k}$ is a $(1-1/e-\varepsilon')$-approximation solution of the $k$-size IM problem, with at least $(1-3n^{-l'})$-probability.
\begin{algorithmic}[1]
\For{$i$ from $1$ to $\log_{2}n-1$}
¡¡¡¡\State Let $c_{i}=(6l\log n+6\log(\log_{2} n))\cdot 2^{i}$;¡¡¡¡
¡¡¡¡\State Let $sum=0$;¡¡¡¡
     \For{$j$ from $1$ to $c_{i}$}
 ¡¡¡¡\State Generate a random RR set $R$;¡¡¡¡
 ¡¡¡¡\State $sum=sum+\kappa(R)$;¡¡¡¡
         \If{$sum/c_{i}>1/2^{i}$}
             \State \Return $KPT^{*}=n\cdot sum/(2\cdot c_{i})$.
          \EndIf
      \EndFor
\EndFor
\State \Return $KPT^{*}=1$.
\State Let $\mathcal{R}'$ be the set of all RR sets generated in the last iteration of the above loop;
\State Innitialize: $S_{k}'=\emptyset$;

\For{$i$ from $1$ to $k$}
¡¡¡¡\State Identify the node $v_{i}$ that covers the most RR sets in $\mathcal{R}'$;¡¡¡¡
¡¡¡¡\State Add $v_{i}$ into $S_{k}'$;¡¡¡¡
¡¡¡¡\State Remove from $\mathcal{R}'$ all RR sets that are covered by $v_{i}$;¡¡¡¡
\EndFor
\State Let $\bar{\lambda}=(2+\bar{\varepsilon})l'n\log n\cdot(\bar{\varepsilon})^{-2}$;
\State Let $\bar{\theta}=\bar{\lambda}/KPT^{*}$;
\State Initialize a set $\mathcal{R}''=\emptyset$;
\State Generate $\bar{\theta}$ random RR sets and put them into $\mathcal{R}''$;
\State Let $\bar{f}$ be the fraction of the RR sets in $\mathcal{R}''$ that is covered by $S_{k}'$;
\State Let $KPT'=\bar{f}\cdot n/(1+\bar{\varepsilon})$;
\State $KPT^{+}=\text{max}\{KPT',KPT^{*}\}$;
\State Let $\lambda'=(8+2\varepsilon)n\cdot(l^{'}\log n+\log 2+\log \tbinom{n}{k})\cdot (\varepsilon^{'})^{-2}$;
\State Let $\theta'=\lambda'/ KPT^{+}$;
\State Initialize a set $\mathcal{R}^{*}=\emptyset$;
\State Generate $\theta'$ random RR sets and insert them into $\mathcal{R}^{*}$;
\State Initialize a node set $S_{k}=\emptyset$;
\For{$i$ from $1$ to $k$}
¡¡¡¡\State Identify the node $v_{i}$ that covers the most RR sets in $\mathcal{R}^{*}$;¡¡¡¡
¡¡¡¡\State Add $v_{i}$ into $S_{k}$;¡¡¡¡
¡¡¡¡\State Remove from $\mathcal{R}^{*}$ all RR sets that are covered by $v_{i}$;¡¡¡¡
\EndFor
\State Let $f$ be the fraction of the RR sets in $\mathcal{R}^{*}$ that is covered by $S_{k}$;
\State \Return $\hat{\sigma}(S_{k})=n\cdot f$.
\end{algorithmic}
\end{algorithm}




\section{References}
\begin{thebibliography}{00}
\bibitem{Goldenberg1} J. Goldenberg, B. Libai, E. Muller, ``Talk of the network: a complex systems look at the underlying process of word-of-mouth,'' in Marketing Letters, 2001, pp. 211--223.
\bibitem{Goldenberg2} J. Goldenberg, B. Libai, E. Muller, ``Using complex systems analysis to advance marketing theory development,'' in
Academy of Marketing Science Review, 2001.
\bibitem{Granovetter} M. Granovetter, ``Threshold models of collective behavior,'' in American Journal of Sociology, 1978, pp. 1420--1443.
\bibitem{Schelling} T. Schelling, ``Micromotives and Macrobehavior,'' in Norton, 1978.
\bibitem{Borodin} A. Borodin, Y. Filmus, and J. Oren, ``Threshold models for competitive inf\mbox{}luence in social networks,'' in WINE, 2010, pp. 539--550.
\bibitem{Chen10} W. Chen, C. Wang, and Y. Wang, `` Scalable inf\mbox{}luence maximization for prevalent viral marketing in large-scale social networks,'' in KDD, 2010, pp. 1029--1038.
\bibitem{Chen09} W. Chen, Y. Wang, and S. Yang, ``Ef\mbox{}f\mbox{}icient inf\mbox{}luence maximization in social networks,'' in KDD, 2009, pp. 199--208.
\bibitem{Chen102} W. Chen, Y. Yuan, and L. Zhang, ``Scalable inf\mbox{}luence maximization in social networks under the linear threshold model,'' in ICDM, 2010, pp. 88--97.
\bibitem{Goyal} A. Goyal, W. Lu and L. V. S. Lakshmanan, ``CELF ++: Optimizing the greedy algorithm for inf\mbox{}luence maximization in social networks,'' in WWW, 2011, pp. 47--48.
\bibitem{Jung} K. Jung, W. Heo and W. Chen, ``IRIE: A scalable inf\mbox{}luence maximization algorithm for independent cascade model and its extensions,'' in ICDM, 2012, pp. 1--20.
\bibitem{Kempe03} D. Kempe, J. M. Kleinberg, and V. Tardos, ``Maximizing the spread of inf\mbox{}luence through a social network,'' in KDD, 2003, pp. 137--146.
\bibitem{Kempe05} D. Kempe, J. M. Kleinberg, and V. Tardos, ``Inf\mbox{}luential nodes in a dif\mbox{}fusion model for social networks,'' in ICALP, 2005, pp. 1127--1138.
\bibitem{Kempe15} D. Kempe, J. M. Kleinberg, and V. Tardos, ``Maximizing the spread of inf\mbox{}luence through a social network,'' in Theory, 2015, pp. 105--147.
\bibitem{Leskovec} J. Leskovec, A. Krause, C. Guestrin, C. Faloutsos, J. VanBriesen and N. Glance, ``Cost-ef\mbox{}fective outbreak detection in networks,'' in KDD, 2007, pp. 420--429.
\bibitem{Leskovec2014} J. Leskovec and A. Krevl, ``SNAP Datasets: Stanford large network dataset collection,'' http://snap.stanford.edu/data, 2014.
\bibitem{Zhang16} H. Y. Zhang, H. L. Zhang, A. Kuhnle, M. T. Thai, ``Prof\mbox{}it maximization for multiple products in online social networks,'' in INFOCOM, 2016, pp. 1--9.
\bibitem{Borgs14} C. Borgs, M. Brautbar, J. T. Chayes, and B. Lucier, ``Maximizing social inf\mbox{}luence in nearly optimal time,'' in SODA, 2014, pp. 946--957.
\bibitem{Sviridenko04} M. Sviridenko, `` A note on maximizing a submodular set function subject to a knapsack constraint,'' in Operations Research Letters, 2004, pp. 41--43.
\bibitem{Tang14} Y. Tang, X. Xiao, and Y. Shi, ``Inf\mbox{}luence maximization: near-optimal time complexity meets practical ef\mbox{}f\mbox{}iciency,'' in SIGMOD, 2014, pp. 75--86.
\bibitem{Tang15} Y. Tang, Y. Shi, and X. Xiao, ``Inf\mbox{}luence maximization in near-linear time: A martingale approach," in SIGMOD, 2015, pp. 1539--1554.
\bibitem{Nguyen16} H. T. Nguyen, M. T. Thai, and T. N. Dinh, ``Stop-and-stare: Optimal sampling algorithms for viral marketing in billion-scale networks," in SIGMOD, 2016, pp. 695--710.
\bibitem{Nguyen13} H. Nguyen and R. Zheng, ``On budgeted inf\mbox{}luence maximization in social networks," in IEEE J. Sel. Area Comm. 31, 2013, pp. 1084--1094.
\bibitem{Du14} N. Du, Y. Liang M. F. Balcan and L. Song, ``Budgeted inf\mbox{}luence maximization for pultiple products," arXiv:1312.2164.
\bibitem{Chernoff}R. Motwani and P. Raghavan, \emph{Randomized Algorithms}, Cambridge University Press, 1995, pp. 68--70.
\bibitem{Hung17} H. T. Nguyen, M. T. Thai, and T. N. Dinh, ``A billion-scale approximation algorithm for maximizing benef\mbox{}it in viral maketing," in IEEE/ACM Transactions on Networking, 2017, pp. 2419--2429.
\bibitem{Moore59} E. F. Moore, ``The shortest path through a maze," in the proceeding of Int. Symp. Switching Theory, 1959, pp. 285--292.
\end{thebibliography}


\end{document}